\documentclass[showpacs,aps,amsmath,amssymb,twocolumn,prb,
nofootinbib]{revtex4}

\usepackage{graphicx}
\usepackage{color}
\usepackage{bbold}

\renewcommand{\vec}[1]{{#1}}
\newcommand{\np}{N_p}
\newcommand{\nomp}{N_{1-p}}
\newcommand{\bit}{\begin{itemize}}
\newcommand{\eit}{\end{itemize}}
\newcommand{\tp}{\mathsf{T}}
 
\newcommand{\f}{\frac}
\renewcommand{\>}{\right\rangle}
\newcommand{\<}{\left\langle}
\newcommand{\ba}{\begin{align}}
\newcommand{\ea}{\end{align}}
\newcommand{\be}{\begin{equation}}
\newcommand{\ee}{\end{equation}}    
\newcommand{\bi}{\begin{itemize}}
\newcommand{\ei}{\end{itemize}}
\newcommand{\lf}{\left(}
\newcommand{\ri}{\right)}
\newcommand{\dd}{\mathrm{d}}

\newcommand{\Tr}{\operatorname{Tr}}
\newcommand{\tr}{\operatorname{tr}}
\newcommand{\str}{\operatorname{str}}

\newcommand{\ret}{

}
\newcommand{\up}{\left| \uparrow \>}
\newcommand{\down}{\left| \downarrow \>}

\newcommand{\cp}{CP}

\renewcommand{\vec}[1]{{\bf #1}}

\newcommand{\Zloop}{Z_{\text{loops}}}
\newcommand{\Lss}{\mathcal{L}_\text{soft}}
\newcommand{\Lsigma}{\mathcal{L}_\text{$\sigma$}}

\newcommand{\uresc}{u}

\begin{document}

\newcommand{\bra}[1]{\< #1 \right|}
\newcommand{\ket}[1]{\left| #1 \>}

\title{Phase transitions in 3D loop models and the $CP^{n-1}$ sigma model}
\author{Adam Nahum and J. T. Chalker}
\affiliation{Theoretical Physics, Oxford University, 1 Keble Road, Oxford OX1 3NP, United Kingdom}
\author{P. Serna, M. Ortu\~no and A. M. Somoza}
\affiliation{Departamento de F\'isica -- CIOyN, Universidad de Murcia, Murcia 30.071, Spain}
\date{September 27, 2013}

\begin{abstract}
We consider the statistical mechanics of a class of models involving close-packed loops with fugacity $n$ on three-dimensional lattices. The models exhibit phases of two types as a coupling constant is varied: in one, all loops are finite, and in the other, some loops are infinitely extended. We show that the loop models are discretisations of  $CP^{n-1}$ $\sigma$ models. The finite and infinite loop phases represent, respectively, disordered and ordered phases of the $\sigma$ model, and we discuss the relationship between loop properties and $\sigma$ model correlators. On large scales, loops are Brownian in an ordered phase and have a non-trivial fractal dimension at a critical point. We simulate the models, finding continuous transitions between the two phases for $n=1,2,3$ and first order transitions for $n\geq 4$. We also give a renormalisation group treatment of the $CP^{n-1}$ model that shows how a continuous transition can survive for values of $n$ larger than (but close to) two, despite the presence of a cubic invariant in the Landau-Ginzburg description. The results we obtain are of broader relevance to a variety of problems, including $SU(n)$ quantum magnets in (2+1) dimensions, Anderson localisation in symmetry class C, and the statistics of random curves in three dimensions. 
\end{abstract}
\maketitle

\suppressfloats

\section{Introduction}


This paper is concerned with the statistical physics of a family of three-dimensional (3D) lattice models for completely-packed loops that have transitions between phases of two types: one in which there are only short loops, and another in which some loops are extended. These models and phase transitions are interesting from several perspectives, since loops play an important role in a variety of problems from classical statistical mechanics and are also central to simulations of quantum systems. Our aim is to identify continuum field theories that describe long-distance properties of the models, to pin down the relation to quantum problems,  and to study the phase transitions using large-scale Monte Carlo simulations. We have previously described some of this work in outline.\cite{prl} 
Here we present a full account as well as additional results.

The models we discuss are characterised by a loop fugacity $n$, which can be interpreted as the number of possible colourings for each loop, and by a coupling constant
$p$, which controls the distribution of loop lengths; full definitions are given in Sec.~\ref{models}.
A defining feature of loop models, in contrast to spin systems, is that the basic degrees of freedom are extended objects. We show in Sec.~\ref{theory}, however, that the partition 
functions can be re-written in terms of local degrees of freedom living on the complex projective space  $CP^{n-1}$ and in this way we identify the loop models as discretisations of $CP^{n-1}$ $\sigma$ models. In addition, a dictionary can be established, expressing correlation 
functions of the $\sigma$ models in terms of those of the loop models: most importantly, the 
two-point correlation function of the $\sigma$ model is related to the probability that two links of the lattice lie on the same loop. The short-loop phase therefore represents the disordered phase of the $\sigma$ model, while infinite loops encode long range order of the $\sigma$ model. The relationship between correlation functions carries implications for the geometry of loops: in particular, the physics of Goldstone fluctuations implies that loops in the ordered phase are Brownian at large distances, while at a critical point loops have a fractal dimension related to the correlation exponent $\eta$ of the $\sigma$ model.

Depending on the value of the loop fugacity, the loop models can be mapped to various problems in critical phenomena. Setting $n=1$, they represent an important class of classical phase transitions which (like percolation) are geometrical 
rather than thermodynamic in nature, in the sense that they are visible only in geometrical observables. In this correspondence, the loops represent line defects in an environment with quenched disorder. Examples include the zero lines of a random complex field,
\cite{tricolour percolation} cosmic strings, \cite{cosmic strings} or optical vortices. \cite{dennis}  The loop models at $n=1$ also arise via exact mappings from network models for certain Anderson metal-insulator transitions. \cite{glr,beamond cardy chalker} At this value of $n$, the field theory must either be construed as a replica limit, or augmented with fermionic degrees of freedom and supersymmetry. 
The properties of such $\sigma$ models and their connection with  loop models
have been discussed recently for two-dimensional (2D) systems \cite{read saleur,candu et al} while in previous work we have studied 3D loop models at $n=1$ \cite{ortuno somoza chalker} and their general relation to geometrical phase transitions. \cite{vortex paper}

Remarkably, when $n$ is greater than one, the same 3D loop models can be mapped to quantum $SU(n)$ antiferromagnets in $(2+1)$ dimensions by considering an appropriate transfer matrix. The phase with long loops is then the Neel phase, and one with only short loops is a valence bond liquid in which spins dimerise without breaking lattice symmetries.  
Indeed, the models we discuss here are closely related to loop algorithms that
have been developed for Monte Carlo simulation of quantum spin 
systems.\cite{QMCreviews,troyer,alet} 
A feature specific to our three-dimensional lattice model is that, since the `time-like' axis is microscopically equivalent to the two `space-like' ones, at a continuous quantum phase transition the 
dynamical exponent value $z=1$ is guaranteed, rather than a matter for calculation. 
Loop models with `deconfined' transitions to valence bond solid phases may also be constructed
by introducing extra couplings, and will be discussed in a separate paper.\cite{interacting}

Besides their appearance in computational algorithms for quantum antiferromagnets, 
loop models with $n>1$ have been examined previously in a variety of other contexts.
The 2D $CP^{n-1}$ sigma model has been simulated 
both by using the connection between $SU(n)$ magnets
in $(1+1)$ dimensions and classical $\sigma$ models in $2$ dimensions,\cite{wiese05} and by means of
a loop representation different from the one we describe here.\cite{wolff strong coupling}
Separately, there have been detailed studies of the properties of loops that
arise in the description of some frustrated classical antiferromagnets,\cite{jaubert,sondhi}
and of cycles that occur in statistical problems involving 
random permutations.\cite{schramm,ueltschi}

Loop models can be studied efficiently using Monte Carlo techniques, and we use simulations
both as a check on our identification of them with  $\sigma$ models and to investigate the phase transitions: results are presented in Sec.~\ref{simulations}. As a test we examine critical phenomena at $n=2$: the $CP^1$  model is known to be equivalent to the ${O}(3)$ (or classical Heisenberg) model, and our results are consistent with previous high precision investigations\cite{O(3) numerics} of that universality class.  
Moving beyond this check, an important basic question concerns the
nature of the transition for general $n$. For $n>2$, Landau theory allows a cubic invariant, implying discontinuous ordering. 
We show, however, in Sec. \ref{fourminusepsilon}, using an expansion around $n=2$ and dimension $d=4$, that 
fluctuations lead to a continuous transition for $n<n_c$, with $n_c >2$ if $d<4$.
At $n=3$ in 3D we find behaviour consistent with a continuous transition, and obtain the exponent values $\nu= 0.536(13)$ and $\gamma = 0.97(2)$. If this is indeed a new critical point, it implies the possibility of similar behaviour in two-dimensional quantum $SU(3)$ magnets. Recent results \cite{kaul-su(n)bilayer} for a bilayer $SU(3)$ magnet are consistent with this. Large loop fugacity favours the disordered phase of loop models, which occupies a growing portion of the phase diagram with increasing $n$. More detailed features vary with the choice of lattice: for one (termed the K-lattice below) we find a first-order transition at $n\geq 4$ between ordered and disordered phases; another (the three-dimensional L-lattice) supports only disordered phases at $n\geq 5$. First order transitions have also 
been reported\cite{duane green,kataoka} from Monte Carlo simulations of other lattice discretisations of the $CP^{n-1}$ model at $n=4$, and from an analytical treatment of the large-$n$ limit.\cite{kunz}



%

\section{Models}\label{models}

The models we study are defined as follows. We start with
a directed lattice of coordination number four
that has two links entering each node and two links leaving. 
A close-packed loop configuration is constructed by selecting
for every node one of the two possible pairings of incoming with outgoing links. 
The statistical weight of such a configuration has two contributions. 
First, for each node a probability $p$ is associated with one pairing of links, and
$1-p$ with the other. Second, each loop carries a fugacity $n$. 
Consider a configuration $\cal C$ in
which the numbers of nodes with each type of pairing are $N_p$ and $N_{1-p}$ respectively,  
and there are $|{\cal C}|$ loops, 
and let the partition function be $Z_{\rm loops}$.
Then
the probability of this configuration is
\begin{equation}
Z_{\rm loops}^{-1}\, p^{N_p} (1-p)^{N_{1-p}} n^{|{\cal C}|}
\end{equation}
and
\begin{equation}
Z_{\rm loops} = \sum_{\cal C} p^{N_p} (1-p)^{N_{1-p}} n^{|{\cal C}|}\,.
\end{equation}
The loop fugacity can be generated by allowing each loop independently to have one of $n$
colours, and summing over loop colours as well as node pairings.
Making an obvious analogy with a Boltzmann weight, we refer to
\begin{eqnarray}
E&\equiv& -N_p \ln p - N_{1-p} \ln(1-p) 
\label{energy}
\end{eqnarray}
as the energy of a configuration.

\begin{figure}[!h] 
\centering
\includegraphics[height=1.4in]{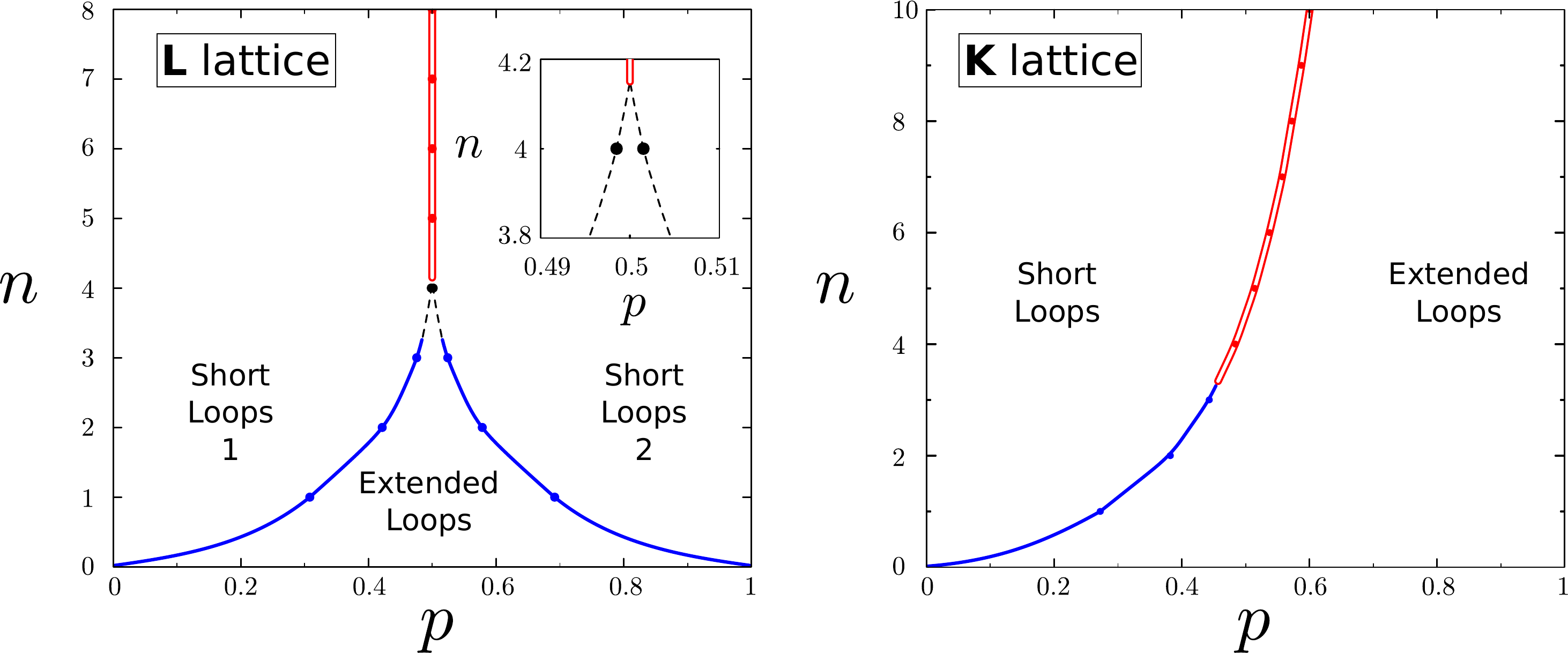} 
\caption{(colour online) Phase diagrams for the L and K lattices. Continuous transitions are indicated by blue dots and a single line and first order transitions by red dots and a double line. For the L lattice, the point $p=1/2,\,n=4$ lies in the extended phase, as shown in the inset.}
\label{phase diagram}
\end{figure}

A model is fully specified by the choices of lattice and link directions, and of which pairing
attracts which weight at each node. We study two models on three-dimensional directed lattices proposed by Cardy.\cite{cardy class C review} They are analogues of the two-dimensional L-lattice and Manhattan lattice and we refer to them as the three-dimensional L-lattice and the K-lattice, respectively. 
The loop model on the three-dimensional L-lattice is symmetric under $p\to 1-p$.
At $p=0$ and $p=1$ it has only loops of minimal length (six steps), but an extended phase
occurs near $p=1/2$ provided $n$ is not too large ($n\leq n^*$, with $4<n^*<5$). 
The loop model on the K-lattice
is not symmetric under $p\to 1-p$; instead, it is designed to ensure that
both localised and extended phases occur as $p$ is varied. 
At $p=0$ it has only loops of minimal length, but at
$p=1$ all trajectories are extended. It has a transition for all $n$ from a localised phase 
at small $p$ to an extended phase near $p=1$. Phase diagrams for loop models on both
lattices are illustrated in Fig.~\ref{phase diagram}.

\begin{figure}[b] 
\centering
\includegraphics[width=1.4in]{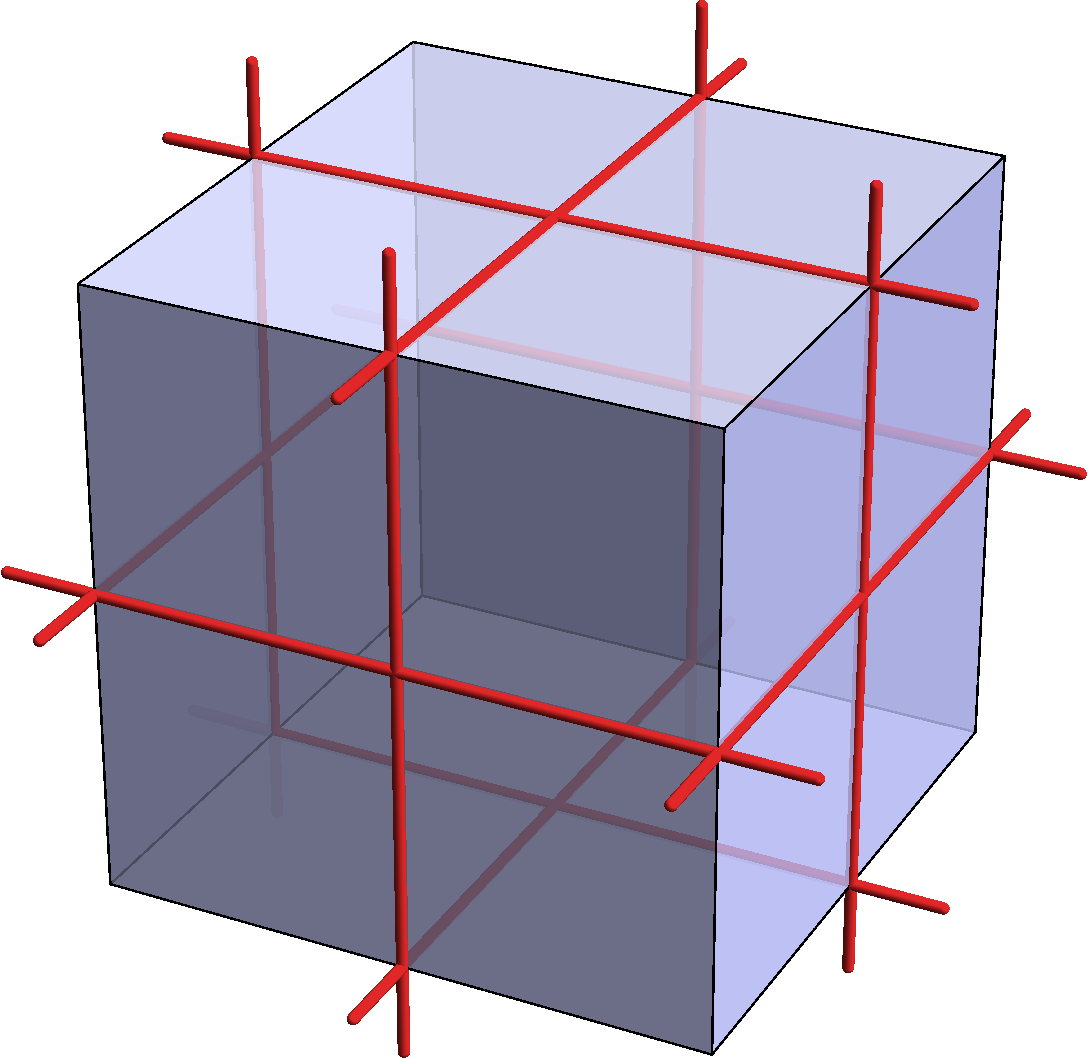}\hspace{0.2in}\includegraphics[width=1.4in]{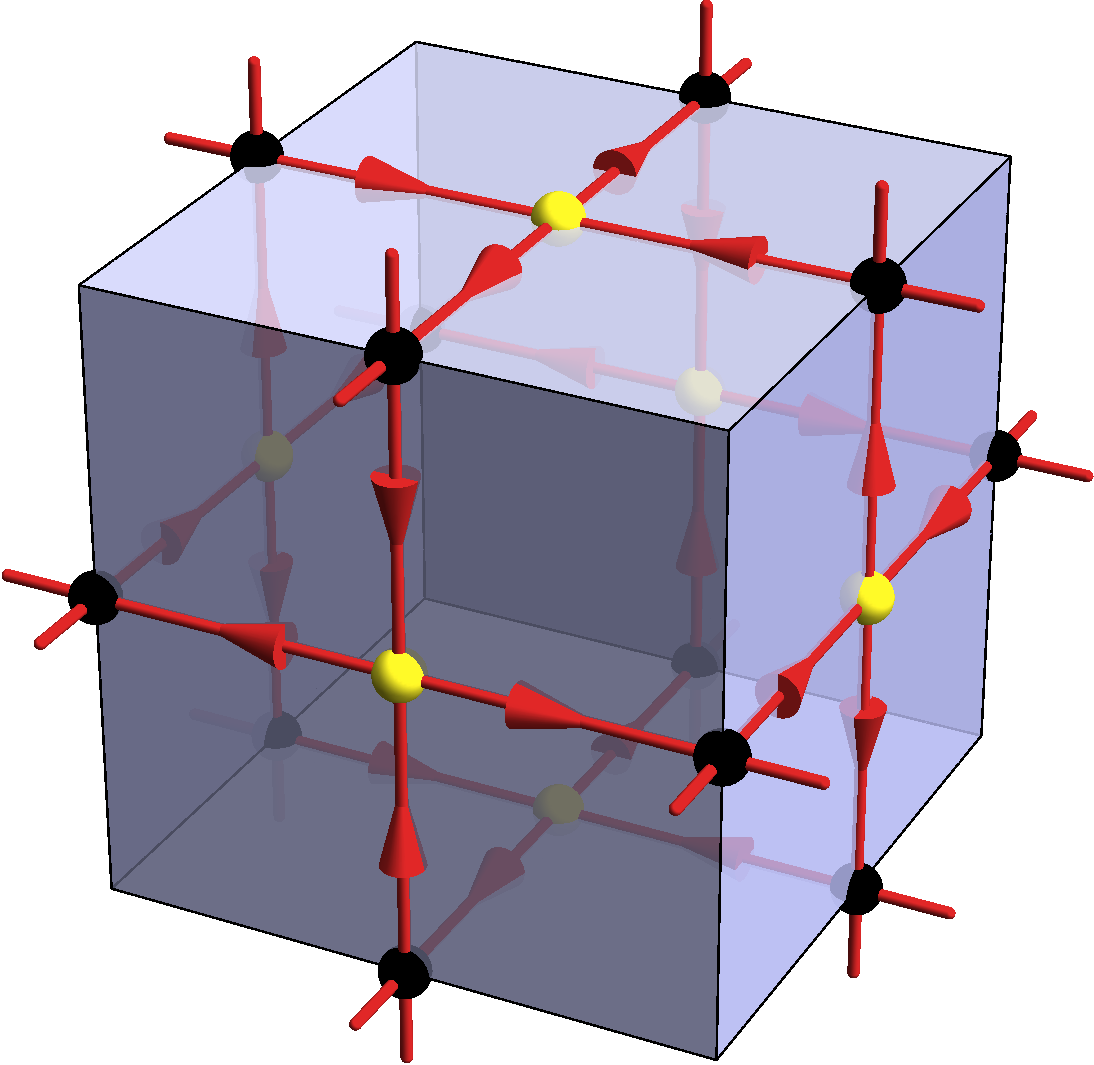}
\caption{(color online) Left: a cube of $C_1$, with the lines of intersection with $C_2$ marked in red. These lines form the links of the $L$ and $K$ lattices. Right: with the orientations corresponding to the $L$ lattice added. The nodes lie on two sublattices, marked in yellow and black.}
\label{C1cubefig}
\end{figure}

Both lattices are defined on a graph $\cal G$ that has cubic symmetry; they differ in their 
link orientations. 
To construct the graph $\cal G$, take two interpenetrating cubic lattices 
$C_1\equiv {(2\mathbf Z)}^3$ and $C_2\equiv (2{\mathbf Z}+1)^3$ as illustrated in Fig.~\ref{C1cubefig}.  The edges (or links) of $\cal G$ are formed by the intersections of the faces of $C_1$ with the faces of $C_2$. The nodes of
$\cal G$ lie on the midpoints of the edges of $C_1$ or of $C_2$ (although these edges do not themselves belong to $\cal G$), and the four links that meet at a node lie on two orthogonal axes.

\begin{figure}[!h] 
\centering
\includegraphics[height=2.4in]{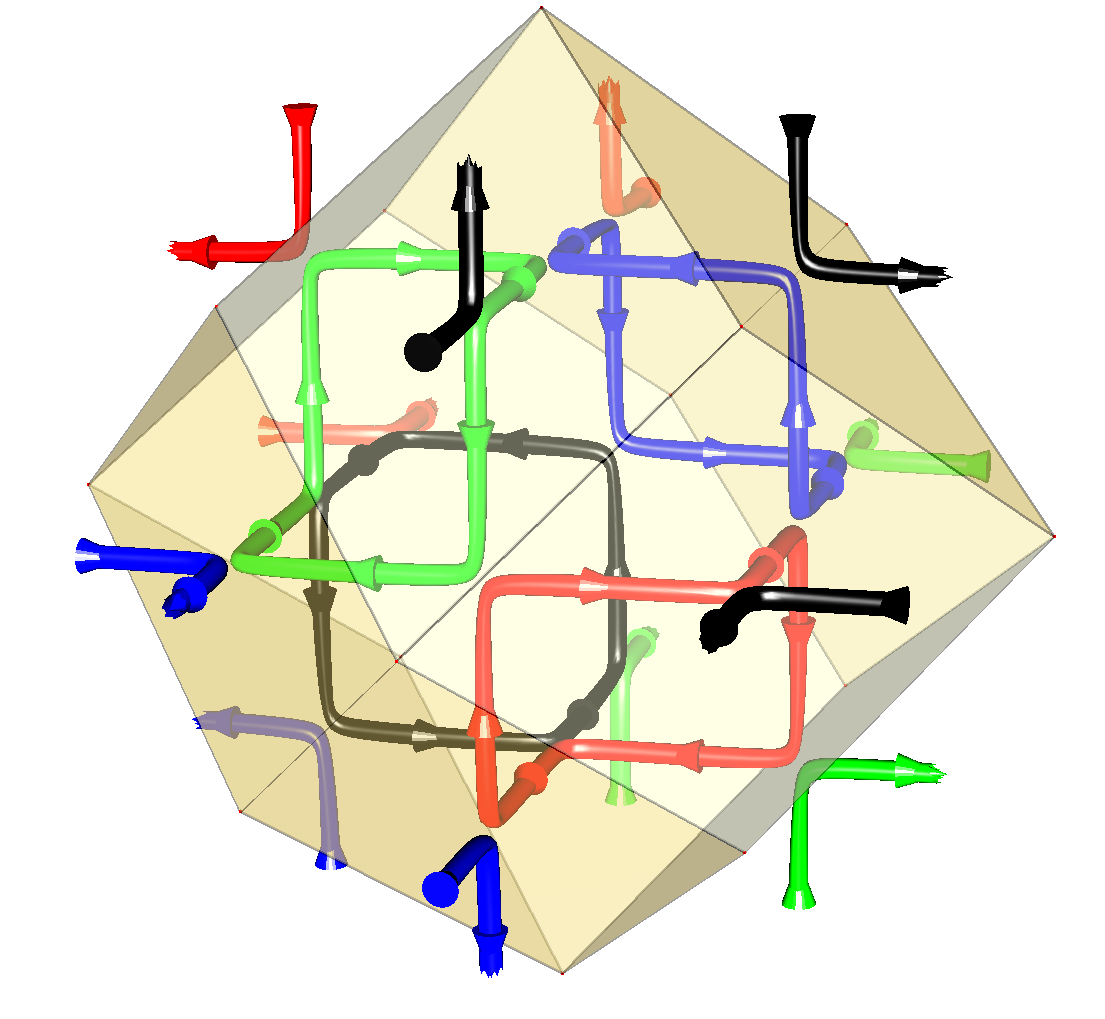} 
\caption{(color online) Loops on the 3D L lattice at $p=0$. The bounding planes of the Wigner Seitz unit cell are shaded and each of the four loops of Table~\ref{hexagons} has a different colour.}
\label{fig1}
\end{figure}

\begin{figure}[h!] 
\centering
\includegraphics[width=3.4in]{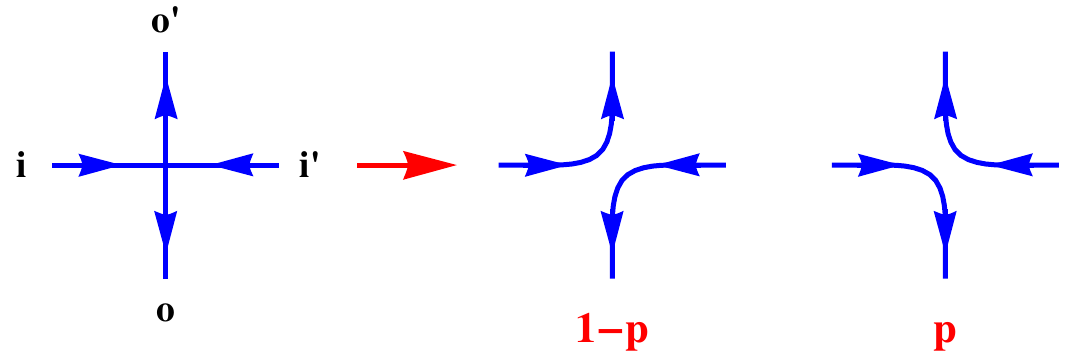} 
\caption{(color online) A node on the L lattice, together with the labelling of links used in Eq.~(\ref{node BW}).}
\label{Llattnode}
\end{figure}

\begin{figure}[h!] 
\centering
\includegraphics[width=3.4in]{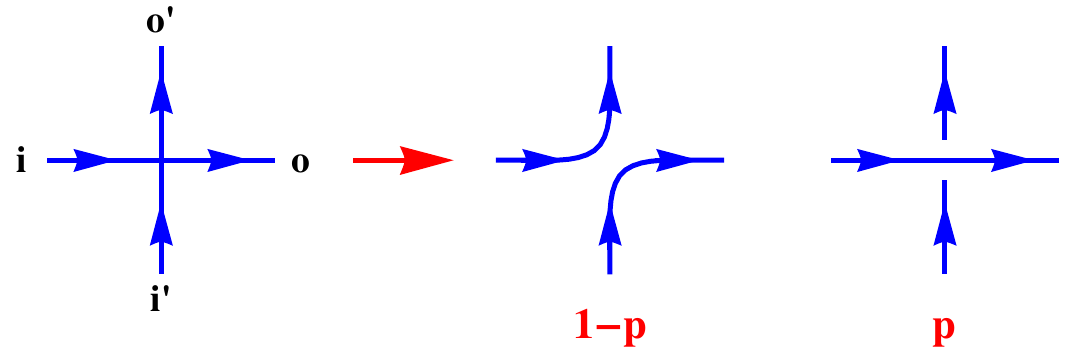} 
\caption{(color online) A node on the K lattice, together with the labelling of links used in Eq.~(\ref{node BW}).}
\label{Klattnode}
\end{figure}

The L-lattice, illustrated in Fig.~\ref{fig1}, has the property that both incoming links at a node lie on the same axis,
and both outgoing links lie on the other axis (Fig.~\ref{Llattnode}). This is sufficient to fix 
the orientation of all links on the lattice up to a  global two-fold choice, which is arbitrary.
The K-lattice 
has the alternative property 
that all links lying on a given axis are directed in the same sense (Fig.~\ref{Klattnode}). In addition, links 
on nearest neighbour parallel axes are oppositely directed.

To specify in a more formal way the assignment of weights $p$ and $1-p$ in each case, we describe the 
unique configuration of loops contributing to $Z_{\rm loops}$ at $p=0$.
For both lattices this consists solely of non-planar loops of six links. 
Each such loop can be defined by giving the coordinates of an initial site
and the orientations of the first three steps from this site, since the remaining 
three steps have opposite orientations in the same order. 
The unit cell of the three-dimensional L-lattice contains four such loops, 
and that of the K-lattice contains two loops, as set out in Table \ref{hexagons}.

We take links to have unit length. 
In simulations we use cubic samples of linear size $L$
with periodic boundary conditions. The number of nodes is then $N=3L^3/4$ for both lattices.


\begin{table}[htbp]
\centering
\begin{tabular}{crrr}
\hline
\hline
\multicolumn{ 2}{c}{L-Lattice} \\ \hline
$(0,1,0)$ & $\hat{\bf x}$&$\hat{\bf y}$&$\hat{\bf z}$\\ \hline
$(1,0,1)$ & $\hat{\bf x}$&$\hat{\bf y}$&$-\hat{\bf z}$\\ \hline
$(0,1,2)$ & $\hat{\bf x}$&$-\hat{\bf y}$&$-\hat{\bf z}$\\ \hline
$(1,2,1)$ & $\hat{\bf x}$&$-\hat{\bf y}$&$\hat{\bf z}$\\ \hline\hline
\end{tabular}\hspace{1cm}
\begin{tabular}{crrr}
\hline
\hline
\multicolumn{ 2}{c}{K-Lattice} \\ \hline
$(0,0,0)$ & $\hat{\bf x}$&$\hat{\bf y}$&$\hat{\bf z}$\\ \hline
$(1,0,0)$ & $\hat{\bf x}$&$-\hat{\bf z}$&$-\hat{\bf y}$\\ \hline\hline
\end{tabular}
\caption{Initial position and first three steps of the hexagons forming the L- and K-lattices.}
\label{hexagons}
\end{table}

\section{Field theory description}\label{theory}

In this section we give a simple way to connect the loop models to field theory (Sections \ref{lattice field theory} and \ref{continuum}). We also discuss (Sec.~\ref{correlators, replica}) the
relationship between loop observables and correlation functions of the field theory, and the
use of a replica-like limit or supersymmetry (SUSY) to extend the range of correlation functions  expressible in that field theory. A compact account of these ideas appeared in Ref.~\onlinecite{prl}; see also the related discussion in Ref.~\onlinecite{vortex paper}.  We discuss the relation to quantum problems described by the same field theories in Sec.~\ref{transfer matrix section}.

\subsection{Introduction of local degrees of freedom}
\label{lattice field theory}

The loop models may be related to lattice `magnets' for spins located on the links $l$ of the lattice. Neglecting for now complications associated with replicas or supersymmetry (and taking $n$ to be a positive integer), these spins are $n$-component complex vectors $\vec z_l$ with
\ba\label{zs}
\vec z_l & = (z^1_l, \ldots, z^n_l) & {\rm and} \qquad
\vec z_l^\dag \vec z_l & = n.
\end{align}
The action for these degrees of freedom will be chosen so that a graphical expansion generates the sum over loop configurations defining $\Zloop$. This leads to a $U(1)$ gauge symmetry,
\be \label{gauge symm}
\vec z_l \sim e^{i\phi_l} \vec z_l,
\ee
implying that the spins live on $\cp^{n-1}$ --- the manifold of fixed-length complex  vectors modulo the equivalence (\ref{gauge symm}). 

The required action may be written as a sum of contributions from nodes. Letting the trace `$\Tr$' stand for an integral over the fixed-length vectors $\vec z$,  normalised so $\Tr 1=1$, we write
\be \label{lattice cpn-1 model}
Z = \Tr  \exp \bigg( - \sum_\text{nodes} S_\text{node} \bigg).
\ee
To define $S_\text{node}$ label the links at a node as in Figs.~\ref{Llattnode}~and~\ref{Klattnode}, with the weight $p$ pairing being $i\rightarrow o$, $i' \rightarrow o'$ and the weight $1-p$ pairing being $i\rightarrow o'$, $i' \rightarrow o$. Then
\be
\label{node BW}
\exp \lf -S_\text{node} \ri = p \,(\vec{z}_o^\dag \vec{z}_i)(\vec{z}_{o'}^\dag \vec{z}_{i'})+(1-p) \, (\vec{z}_{o}^\dag \vec{z}_{i'})(\vec{z}_{o'}^\dag \vec{z}_i).
\ee
The total action is invariant under the gauge transformation (\ref{gauge symm}), as the phase $\phi_l$ cancels between the terms for the two nodes adjacent to link $l$.

The two terms in $e^{-S_\text{node}}$ are in correspondence with the two node configurations in the loop model. This leads to a simple expression for the partition function in terms of loop configurations $\mathcal{C}$, of the form
\be \label{BW expansion}
\Tr \prod_\text{nodes} \exp\lf-S_\text{node}\ri =
\sum_\mathcal{C} \, W_\mathcal{C} 
\hspace{-1.2mm}
\prod_{\text{loops in } \mathcal{C}}  \hspace{-0.2mm}
\Tr \,
(\vec z_1^\dag \vec z^{\phantom{\dag}}_2) \ldots ( \vec z_\ell^\dag \vec z^{\phantom{\dag}}_1).
\ee 
Here we have used the labels $1,\ldots, \ell$ for the links lying on a loop of length $\ell$ and $W_{\mathcal C}=p^{N_p} (1-p)^{N_{1-p}}$ is the weight associated with the nodes. Next, perform the integrals over the $\vec z$s using $\Tr z_l^\alpha \bar z_l^\beta = \delta^{\alpha\beta}$ (repeated indices are summed throughout) to obtain
\ba\notag
\Tr\,(\vec z_1^\dag \vec z^{\phantom{\dag}}_2) \ldots ( \vec z_\ell^\dag \vec z^{\phantom{\dag}}_1)& =
 \Tr \, (\bar z_1^{\alpha_1} z_2^{\alpha_1}) (\bar z_2^{\alpha_2} z_3^{\alpha_2}) \ldots (\bar z_\ell^{\alpha_\ell} z_1^{\alpha_\ell})\\ 
& =  \delta^{\alpha_1 \alpha_2} \delta^{\alpha_2 \alpha_3} \ldots \delta^{\alpha_\ell \alpha_1}.\label{integrals on loop}
\end{align}
The Kronecker deltas force the spin indices $\alpha_i$ to be equal for all the links $i$ on a given loop. The remaining free index gives the desired sum over $n$ colours for each loop, so that
\ba
Z = 
\sum_\mathcal{C} 
\sum_{\substack{\text{loop}\\ \text{colours}}} W_\mathcal{C} = \Zloop.
\end{align}
This establishes the correspondence between the loop models and models with local `magnetic' degrees of freedom. Its utility is that we may now make the simplest conjectures for the models' continuum descriptions, taking account of the $SU(n)$ global and $U(1)$ gauge symmetries of the lattice action appearing in Eq.~(\ref{lattice cpn-1 model}). 

Note that this lattice action is complex-valued. 
When coarse-graining is considered carefully, the naive real action resulting
from a derivative expansion has to be supplemented with imaginary terms associated with hedgehog defects,\cite{interacting} but for the transitions considered in this paper the consequence of these terms is only to renormalise the parameters in the effective Lagrangian, and not to change the naive result.


\subsection{Continuum limit}
\label{continuum}

Let us exchange the vector $\vec z$, which is a redundant parametrisation of $CP^{n-1}$, for the gauge-invariant matrix $Q$, defined by
\be
Q^{\alpha \beta} = z^\alpha \bar z^\beta - \delta^{\alpha \beta}.
\ee
$Q$ is Hermitian and  traceless, and obeys the nonlinear constraint $(Q+1)^2 = n \, (Q+1)$. 

In the continuum we may either retain this constraint, giving the $CP^{n-1}$ $\sigma$ model 
\ba\label{sigma model lagrangian}
\Lsigma & = \f{1}{2 g} \tr \, (\nabla Q)^2 &
&(\text{and constraint on $Q$}),
\end{align}
or we may use a formulation in which $Q$ is an arbitrary traceless Hermitian matrix, giving the soft spin model
\be \label{soft spin lagrangian}
\Lss = \tr \lf\nabla Q\ri^2 + t \tr Q^2 + g \tr Q^3 + \lambda \tr Q^4 + \lambda' (\tr Q^2)^2.
\ee
The existence of hedgehog defects (possible because of the nontrivial second homotopy group $\pi_2(CP^{n-1})=\mathbb{Z}$) means that the second formulation is arguably more natural in three dimensions. Hedgehogs are known to play an important role in the vicinity of the critical point, and they proliferate in disordered phase; they are of course irrelevant in the ordered phase.\cite{Kamal Murthy, Motrunich Vishwanath, senthil et al dcp, quantum criticality beyond, Read Sachdev spin peierls, Haldane O(3) sigma model} In order to accommodate them in the sigma model, the constraint must be relaxed in the defect core, and the regularisation-dependent physics in the core then   determines a finite fugacity for defects --- the condensed formulation (\ref{sigma model lagrangian}), with only the single parameter $g$, is therefore slightly misleading.



The manifold $\cp^1$ is simply the sphere, so at the special value $n=2$ the above field theories reduce to the $\sigma$ model and soft spin incarnations of the $O(3)$ model. 
At this value of $n$, the cubic term in $\Lss$ vanishes, and there is only one quartic term. 
The $O(3)$ spin $S$ is related to $Q$ via the Pauli matrices. Setting $Q = \f{1}{\sqrt{2}}\sigma^i S^i$ in $\Lss$ gives 
\be\label{O(3) Lagrangian}
\Lss = (\nabla S)^2 + t\, S^2 + u \, (S^2)^2
\ee
with $u=\lambda' + \lambda/2$. 

This description leads us to expect continuous transitions in the $O(3)$ universality class for the $n=2$ models. Continuous transitions are also expected for $n<2$. For $n>2$, the naive expectation is a first order transition, as a result of the cubic term in $\Lss$. However, as we will see in Sec.~\ref{fourminusepsilon}, fluctuations can invalidate this mean field prediction when the spatial dimension is less than four, and so numerical work is required to determine what happens in 3D.

\subsubsection{Aside: compact and non-compact $CP^{n-1}$ models}

The field theories described above should be distinguished from the related `non-compact' $CP^{n-1}$ models, in which $\vec z$ is coupled to a non-compact $U(1)$ gauge field $A$:
\be\notag
\mathcal{L}_{\text{NCCP}^{n-1}} = \f{1}{2} |(\nabla-i A)\vec z|^2 + \kappa (\nabla \times A)^2 + \mu |\vec z|^2 + \lambda |\vec z|^4.
\ee
The universal behaviour of the `compact' $CP^{n-1}$ models (those discussed in this paper) may also be captured by gauge theories similar to the above, but with a compact $U(1)$ gauge symmetry. This distinction is discussed in Refs.~\onlinecite{senthil et al dcp}, \onlinecite{quantum criticality beyond} and \onlinecite{Motrunich Vishwanath}. In the compact case, configurations of the gauge field $A$ are allowed that contain Dirac monopoles. This leads to confinement of $\vec z$ quanta, so that at long distances only the neutral degrees of freedom in $Q$ play a role, and we return to the field theories already discussed. 

In our microscopic models, the gauge symmetry is compact: the parameter $\phi_l$ in (\ref{gauge symm}) is defined only modulo $2\pi$. A subtlety is that in some cases \cite{senthil et al dcp,quantum criticality beyond} critical points in models with compact gauge symmetry may be described by emergent non-compact gauge theories in the continuum. This occurs as a result of the suppression of Dirac monopoles --- or, as it turns out equivalently, of hedgehog configurations in $Q$. 

\subsection{Correlation functions}
\label{correlators, replica}

The graphical expansion of the lattice field theory (\ref{lattice cpn-1 model}) generates a sum over loop configurations in which each loop carries a colour index $\alpha = 1,\ldots, n$. This leads straightforwardly to expressions for correlation functions of $Q$ in terms of the probabilities of  geometrically-defined events. These relationships are those one would expect from viewing the loops as worldlines of $\vec z$ quanta, with one of the spatial directions taken as the imaginary time direction for a quantum problem. These worldlines come in $n$ colours, one for each component of $\vec z$, and since $z^\alpha$ is a complex field they carry an orientation distinguishing particles from antiparticles.

In particular, the operator $Q_l^{\mu \nu} = z_l^\mu \bar z_l^\nu$ with $\mu \neq \nu$ absorbs an incoming strand/worldline of colour $\mu$ and emits an outgoing one of colour $\nu$. More precisely, the effect of this operator on the graphical expansion is to force the loop passing through  link $l$ to change colour from $\mu$ to $\nu$ there. This follows from the fact that the 
presence of the operator changes the
single link integral appearing in Eq.~\ref{integrals on loop} from $\Tr z_l^\alpha \bar z_l^{\alpha'} = \delta^{\alpha\alpha'}$ to $\Tr z_l^\alpha \bar z_l^{\alpha'} Q_l^{\mu\nu} = A \, \delta^{\alpha \nu} \delta^{\alpha' \mu}$ (for $\mu\neq\nu$), with $A=n/(n+1)$. 

We can use $Q$ to represent the probability $G(l,l')$ that two links lie on the same loop, since the correlator $\<Q^{12}_l Q^{21}_{l'} \>$ receives contributions only from configurations in which $l$ and $l'$ are joined (by a loop with one arm of colour $1$ and one arm of colour $2$) and
\be\label{15}
G(l,l') = \f{n}{A^{2}} \big\langle Q^{12}_l Q^{21}_{l'} \big\rangle.
\ee
In this expression
the factor of $n$ compensates for the fact that there is no sum over colour indices for the loop passing through $l_1$ and $l_2$. 
In the terminology of loop models, $Q^{\alpha\beta}$ is a `two-leg' operator. 

For convenience, we here use the off-diagonal elements of $Q$ to write geometrical correlators, but all $Q$ correlators can be expressed in terms of loops; for example $\< \tr Q_l Q_{l'} \>=(n-1)A \, G(l,l')$. We may think of the diagonal components, e.g.  $Q^{11}_l$, as operators which measure the colour of a link.


The two-leg correlator $G(l,l')$ generalises to  $2k$-leg correlators $G_{2k}$,  which give the probability that two  regions are joined by $2k$ strands. For example, on the lattice we can define $G_4$ as the probability that four separate strands connect two nodes. In the continuum, such correlators may be written
\be\label{2klegop}
G_{2k}(x, y) \propto \< [Q^{12}(x)]^k [Q^{21}(y)]^k \>,
\ee
where $Q(x)$ is the continuum field.

So far, the correlators we have considered involve only two distinct spin indices $\alpha = 1,2$. They can therefore be written down so long as the spin $\vec z$ has at least two components ($n\geq 2$). More complex correlation functions may require the use of more indices. For example, if $n\geq k$ we can express the probability that a single loop passes through all the links $l_1, \ldots, l_k$, in that order as
\be 
\label{multi-link correlator}
G(l_1, \ldots, l_k) = \f{n}{A^k} \< Q^{12}_{l_1} Q^{23}_{l_2} \ldots Q^{k1}_{l_k} \>.
\ee

\subsubsection{Replicas / SUSY}

These formulas highlight a potential problem. The complexity of the geometrical correlation functions we can represent is limited by the number $n$ of spin components at our disposal, which in turn is set by the loop fugacity. The problem is most acute at $n=1$, when we cannot represent any correlation functions at all. 

The simplest way to get around this is to calculate the desired correlation function assuming that $n$ is sufficiently large, and subsequently analytically continue to the required value of $n$. This is a standard idea in the study of geometrical problems such as loop models and percolation, and is analogous to the replica trick in disordered systems. For simplicity, we will use this replica-like limiting procedure in this paper. However it should be noted that there is also a more rigorous alternative, which is to augment the field theory with additional fermionic degrees of freedom in such a way that the resulting field theory has a global supersymmetry. For our purposes in this paper, replicas and SUSY are equivalent and it is easy to translate between them.

The supersymmetric construction is described in Refs.~\onlinecite{candu et al, read saleur, prl} and \onlinecite{vortex paper}, and leads to the so-called $CP^{n+k-1|k}$ model, in which $\vec z$ is replaced by a supervector $\vec \psi$ with $n+k$ bosonic components $z^\alpha$ and $k$ fermionic ones $\chi^\alpha$, so that
\be
\psi = (z^1, \ldots, z^{n+k}, \chi^1, \ldots, \chi^k).
\ee
The supersymmetry ensures that the value of $k$ does not affect the partition function or its  expansion in loop configurations. However, increasing $k$ yields more operators.


\subsubsection{Phases of the $CP^{n-1}$ model}

When the $CP^{n-1}$ spins are disordered, correlators such as $G(x,y)$ decay exponentially, indicating that long loops are exponentially suppressed. At a critical point, the anomalous dimension $\eta$ of the spin $Q$ determines the decay of $G(x,y)$,
\be
G(x,y) \sim |x-y|^{- (1+\eta)}, 
\ee 
the fractal dimension of a loop \cite{duplantier saleur percolation, scaling relations}
\be
d_f = \f{5-\eta}{2},
\ee
and the probability $P(l)$ that the loop passing though a given link is of length $l$:
\ba
P(l) & \sim l^{1-\tau}, &
\tau & = \f{11 - \eta}{5-\eta}.
\end{align}
If the mean field estimate $\eta=0$ is a good approximation, the loops have a fractal dimension close to $5/2$.

When the spins are ordered, $G(x,y)$ is long-ranged, indicating the appearance of extended loops. The nature of these extended loops in a large, finite sample of linear size $L$ depends on whether curves are allowed to terminate at the boundary or whether all loops are closed. In the former case the extended loops have length $O(L^2)$, since the loops in this phase have fractal dimension two. In the latter case they have a length $O(L^3)$: although the fractal dimension of a segment of radius $\lesssim L$ is still two, the number of times a given extended loop crosses the sample is of order $L$.  We now focus  on the case with only closed loops, which pertains to our simulations with periodic boundary conditions.

The magnitude of order parameter in this phase is proportional to the fraction of links lying on extended loops. 
To demonstrate this, consider applying a magnetic field $h \sim {\cal O}(L^{-2})$ that couples to $Q$ via a term in the action $\delta S = - h \sum_l Q^{11}_l$. In the spin language, the field fixes the direction of symmetry breaking to $\vec z \sim (1, 0, \ldots, 0)$ and the
order parameter is
simply the average $\langle Q^{11}_l\rangle$ calculated in the presence of the weak field $h$. 
In the graphical expansion, $\delta S$ endows loops of colour $\alpha=1$ with a small additional length fugacity: finite strands are insensitive to $h$ but extended strands are forced to be of colour $\alpha=1$, so that they alone contribute to $\langle Q^{11}_l\rangle$.


In the extended phase, contributions to the $2k$-leg correlation functions $G_{2k}$ may be split up according to how many of the strands are finite and how many are infinite, and the power law decay of the geometrical correlation function depends only on the former. Let  $\widetilde G_m$ be the probability that $m$ finite strands connect $x$ and $y$, irrespective of the number of infinite strands. Then we find
\be\label{G finite}
\widetilde G_m(x,y) \sim |x-y|^{-m}.
\ee
These exponents are also independent of how many of the finite strands are oriented from $x$ to $y$ (rather than the reverse) and whether the strands join together form a single loop or many loops. The exponent values  indicate that within the extended phase a single  walk is Brownian on long scales, having for example a fractal dimension of two.

To arrive at this result,  consider again the effect of a weak symmetry breaking field, with
$\delta S = - h \sum_l Q^{11}_l$. 
We can then regard the operators  $z^\alpha$ and $\bar z^\alpha$  for $\alpha>1$ as absorbing/emitting strands which must be finite, while $ z^1$ and $\bar z^1$ absorb/emit strands that may be extended. At large separations, the dominant contribution to any correlator is given by setting $z^1$ equal to a constant (determined by the strength of long range order) while contributions from components with $\alpha >1$ are proportional to Goldstone mode correlators, with two-point functions that decay inversely with separation.  For example the correlator $\< ( z^1 \bar z^2)^m (x)   ( \bar z^1 z^2)^m (y) \>$, which forces $m$ finite strands to propagate from $x$ to $y$, decays as $|x-y|^{-m}$. This argument is readily generalised to give Eq.~(\ref{G finite}).



\subsection{Transfer matrices and quantum magnets}
\label{transfer matrix section}

The link colours provide a convenient basis for the transfer matrix, which acts between `time' slices formed by planes of the lattice and which defines an associated quantum Hamiltonian. We briefly review a standard simple example,\cite{afflecknew} then describe the qualitative features of the Hamiltonians corresponding to the 3D loop models.

\begin{figure}[b] 
\centering
\includegraphics[width=3.2in]{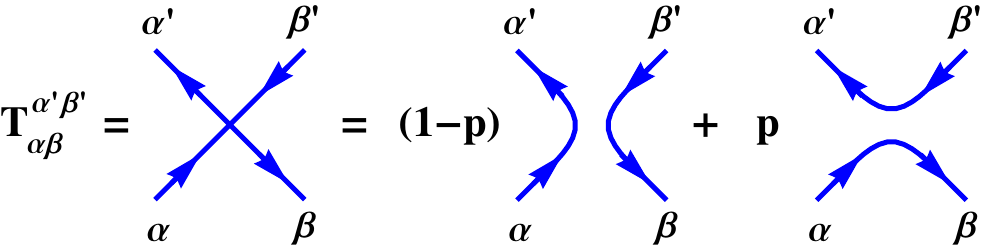} 
\caption{(color online) Graphical representation of the transfer matrix for a single node. The indices $\alpha,\beta,\alpha',\beta'$ denote link colours. Imaginary time increases in the vertical direction, with the lower links being at time $\tau$ and the upper links at time $\tau+\Delta\tau$.}
\label{transfermatrixgraphic}
\end{figure}

Consider first of all the transfer matrix $T$ for a single node of the type shown in Fig.~\ref{transfermatrixgraphic}. In component form this is $T^{\alpha'\beta'}_{\alpha\beta}$, where the upper indices are the colours  of the links at time $\tau + \Delta \tau$, and the lower indices those at time $\tau$. This matrix has two terms
\be \label{single node TM}
T^{\alpha'\beta'}_{\alpha\beta} = (1-p) \delta^{\alpha'}_\alpha \delta^{\beta'}_\beta + p \, \delta^{\alpha'\beta'} \delta_{\alpha\beta}
\ee
corresponding to the two node configurations.
The partition function $\tr T^N$ defines a quasi-one-dimensional loop model with $N$ nodes. The transfer matrix is also the imaginary time evolution operator for a two-site quantum problem with Hamiltonian $H$, related via
\be
T = e^{-\Delta \tau H}.
\ee
Each site (we label them $A$ and $B$) has an $n$-dimensional Hilbert space, spanned by kets $\ket{\alpha}_A$ and $\ket{ \alpha}_B$ respectively. In terms of these
\ba
T & = (1-p) \mathbb{1} + p \sum_{\alpha \beta} \ket{\alpha}_A \ket{\alpha}_B \bra{\beta}_A\bra{\beta}_B \\ 
 & = (1-p)\mathbb{1} +  p\, n \, \mathcal{P}_{AB}.
\end{align}
In the second line $\mathcal{P}_{AB}$ is the projector onto the state $\f{1}{\sqrt{n}} \sum_\alpha \ket{\alpha}_A \ket{\alpha}_B$. This state is a singlet under $SU(n)$ transformations that act as
\ba
\ket{\alpha}_A & \longrightarrow U_{\alpha\beta} \ket{\beta}_A  & {\rm and} \quad
\ket{ \alpha}_B & \longrightarrow U^*_{\alpha\beta} \ket{\beta}_B.
\end{align}
The degrees of freedom at $A$ and $B$ are $SU(n)$ spins, transforming in the fundamental and antifundamental representation respectively.

At $n=2$ they are standard spin-1/2 degrees of freedom. In this case it is convenient to relabel the basis states as $S_z$ eigenstates for spin operators $\vec S_A$ and $\vec S_B$,
\ba\notag
\ket{1}_A& =\up_A, &    \ket{{1}}_B& = \down_B, \\
 \ket{2}_A & = \down_A, & \ket{{2}}_B&=-\up_B.
\end{align}
Both spins then transform in the same representation (this is possible at $n=2$ because  the fundamental representation of $SU(2)$ is pseudoreal). The singlet takes the usual form $\f{1}{\sqrt 2} [ \up_A \down_B - \down_A \up_B ]$, and the projector onto it is $\mathcal{P}_{AB} =  1/4 - \vec S_A. \vec S_B$.

For the single node we can easily extract the form of the Hamiltonian. 
Dropping an additive constant, it is
\be
H = J \, \mathcal{P}_{AB},
\ee
with $J = \Delta\tau^{-1} \ln[1+pn/(1-p)]$. 

\begin{figure}[t] 
\centering
\includegraphics[width=3.2in]{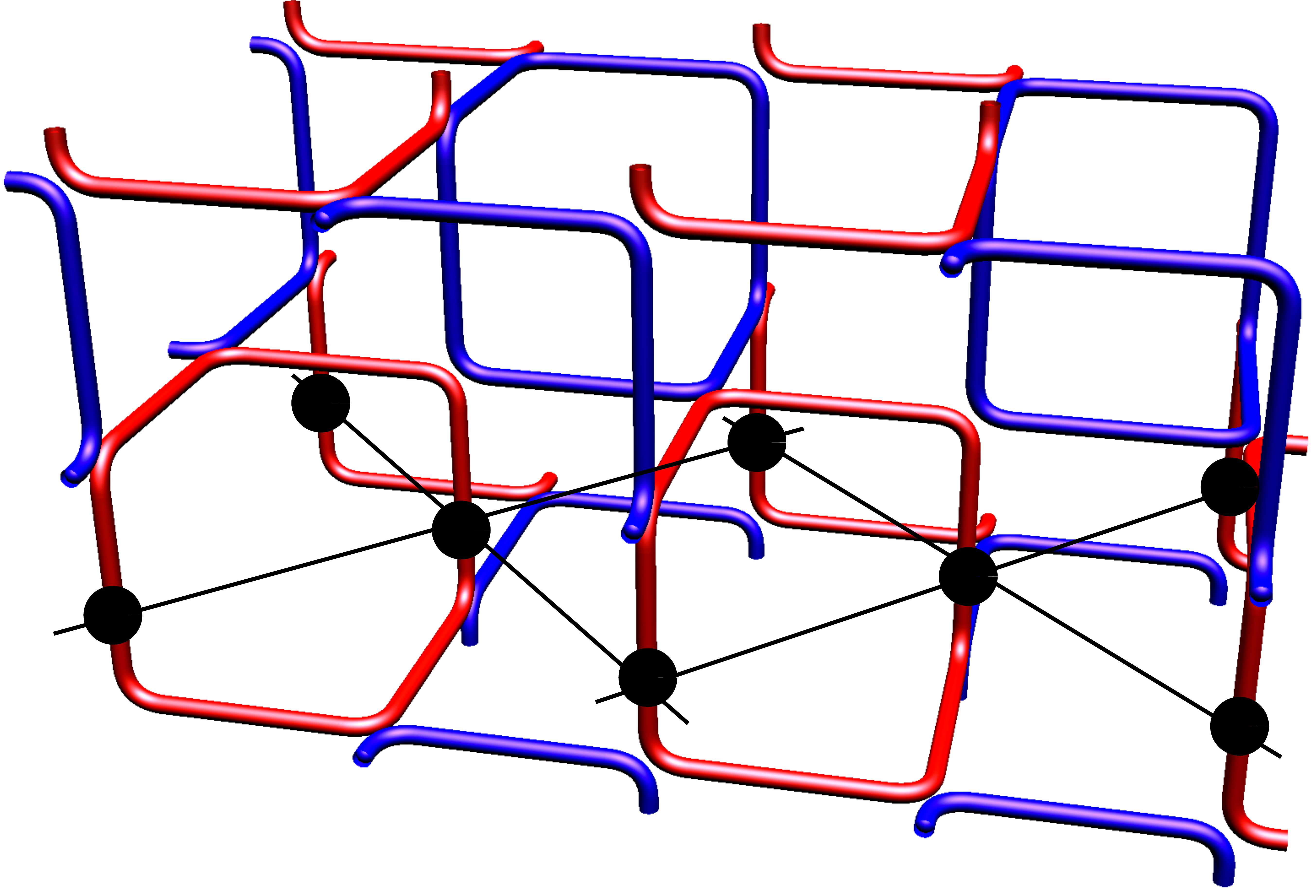} 
\caption{(color online) The K lattice at $p=0$: this short loop phase corresponds to a staggered dimer state. The square lattice for the associated quantum problem is formed by the links and filled circles in a time slice, shown in black.}
\label{transfermatrixillustration}
\end{figure} 

For 2D or 3D loop models, the Hamiltonian will not generally take a simple explicit form. If we insist on one, we must take a continuum limit in imaginary time --- this corresponds to making the node weights anisotropic in such a way that the transfer matrix becomes close to the identity. This procedure is standard for the loop model on the two-dimensional L lattice:\cite{afflecknew}  the resulting Hamiltonian describes a nearest-neighbour spin chain, and the node parameter in the loop model controls dimerisation in the strength of the exchange. For the 3D models we take the view that the precise form of $H$ is less important than the degrees of freedom, symmetries and phase structure of $H$.

For both the L and K lattice we take imaginary time to run parallel to the vertical axis of Fig.~\ref{C1cubefig}. The links intersecting a time slice (which is of thickness two link lengths in the K lattice and four in the L lattice) then form a square lattice with lattice spacing $\sqrt{2}$, as shown in Fig.~\ref{transfermatrixillustration}. One sublattice consists of upgoing and the other of downgoing links; as in the single-node example, this leads to an $SU(n)$ magnet with spins in the fundamental representation on one sublattice and in the antifundamental representation on the other. The transfer matrix $T$ for a given lattice is a sum over configurations within the timeslice, in analogy to Eq.~(\ref{single node TM}). The phase structure of the models is like that of nearest-neighbour $SU(n)$ magnets with dimerization in the strength of the exchange, $H = \sum_{\<ij\>} J_{ij} \mathcal{P}_{ij}$; of course the actual Hamiltonian, given by the logarithm of $T$, would not take this simple form.

The extended phase in the loop models corresponds to the N\'eel phase, while the short-loop phase corresponds to a dimerised phase. The pattern of dimerisation in a short-loop phase can be seen from the representative configuration in which all loops have the minimal length of six. These loops connect the links within a timeslice in pairs; in the quantum problem singlets form between the paired sites. For the L lattice there are four packings of minimal-length loops,  corresponding to the four columnar packings of singlets on the square lattice. When $p=1/2$ these packings are related by lattice symmetry, which is broken when $p\neq 1/2$. That is, varying $p$ away from $1/2$ imposes a specific dimerisation in the couplings in $H$. For the K lattice there is only a single packing of minimal-length loops, which corresponds to a staggered packing of singlets.

\section{The $CP^{n-1}$ model near $n=2$ and $d=4$}
\label{fourminusepsilon}

At $n=2$, when the cubic term in $\Lss$ vanishes, the upper critical dimension of the $CP^{n-1}$ model is four rather than six.
This allows a double expansion in
\be
\Delta = n-2 \qquad\quad \text{and} \qquad\quad \epsilon = 4 - d.
\ee
This idea has been discussed previously for the $Q$-state Potts model, where the expansion is about the Ising limit.\cite{Newman Riedel Muto} The conclusions below for the $\cp^{n-1}$ model are qualitatively identical. In particular, a universal $n_c$ appears, which is greater than the mean field value two when $d<4$. 

To begin with, recall the Wilson-Fisher \cite{Wilson Fisher} renormalisation group (RG) equations for the $O(3)$ (or $CP^1$) model  (\ref{O(3) Lagrangian}). To lowest nontrivial order these are, after rescaling $u\rightarrow\uresc /22$,
\ba \label{Wilson Fisher RG equations}
\f{\dd \uresc}{\dd \ln L} & = \epsilon \uresc - \uresc^2, &
\f{\dd t}{\dd \ln L} & = \lf 2 - \f{5}{11} \uresc \ri t.
\end{align}
Setting $t=0$, there are fixed points at $\uresc=0$ and $\uresc=\epsilon$. For $\epsilon>0$, the latter is stable in the $\uresc$ direction, and describes the critical $CP^1$ model, while the former is unstable in the $\uresc$ direction and describes the tricritical point.

We now consider a formal expansion of these equations in $\Delta$ (compare the approach to the $2+\epsilon$ dimensional $O(n)$ model in Ref.~\onlinecite{Cardy Hamber}), deferring the field-theoretic interpretation until Sec.~\ref{more concrete picture}. The leading contribution is a modification to the RG equation for $u$:
\ba \label{modified RG equation}
\f{\dd \uresc}{\dd \ln L} & = - a \Delta + \epsilon \uresc - \uresc^2.
\end{align}
The consistent scaling is to take $\Delta$ to be $O(\epsilon^2)$, and $\uresc$ to be $O(\epsilon)$ as at the Wilson Fisher fixed point. Here $a$ is an undetermined universal constant, assumed positive in order to give sensible RG flows.

\begin{figure}[t] 
\centering
\includegraphics[height=1.25in]{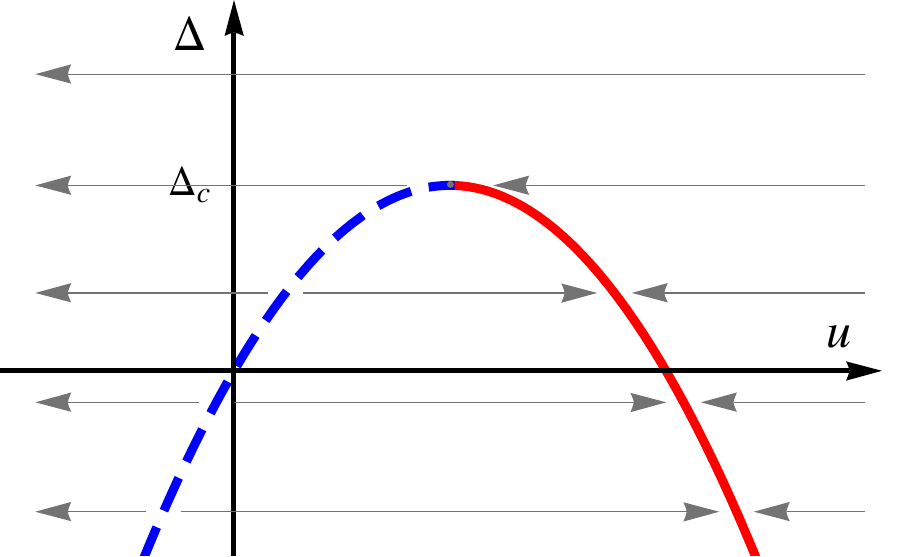}
\caption{(color online) RG fixed points and flows in the $(u,\Delta)$ plane for $d<4$, showing critical (red full curve) and tricritical (blue dashed curve) fixed points merging at $\Delta_c$.}
\label{RGflowsdlessthanfour}
\end{figure}

Fig.~\ref{RGflowsdlessthanfour} shows the resulting RG fixed points in the $(\uresc, \Delta)$ plane for fixed $\epsilon>0$. As $\Delta$ is increased from zero, the critical and tricritical fixed points approach each other, annihilating at a critical $n_c$ given to this order by 
\ba
n_c \simeq 2 + \f{\epsilon^2}{4 a}.
\end{align}
The thermal and leading irrelevant exponents at the critical point are
\ba \notag
y_t &\simeq 2 - \f{5}{22} \lf \epsilon + \sqrt{\epsilon^2 - 4 a \Delta} \ri, & 
y_\text{irr} & \simeq - \sqrt{\epsilon^2 - 4 a \Delta}.
\end{align}
The anomalous dimension $\eta$ is $O(\epsilon^2)$, as in the $O(N)$ model. Analogous formulas hold for the Potts model.\cite{Newman Riedel Muto}

Precisely at $n_c$, the irrelevant exponent vanishes and there are logarithmic corrections to scaling. 
Above $n_c$, there are no fixed points: the RG flows go off to large negative $u$, which we interpret  as a first order transition. 


The size of discontinuities at this first order transition decrease rapidly as $n$ approaches $n_c$ from above. Integrating Eq.~(\ref{modified RG equation}) from a microscopic scale at which $\uresc$ is positive and of order one to the scale of the correlation length $\xi$, where $\uresc$ is negative and of order one, gives
\be\label{ncrit}
\xi \sim  \exp \lf \f{2\pi}{\epsilon} \sqrt{\f{n_c - 2}{n-n_c}} \ri.
\ee
Similar forms hold for other quantities such as the latent heat.\cite{cardy nauenberg scalapino} The asymptotic form $\xi \sim \exp \lf \text{const.}/ \sqrt{n-n_c} \ri$ is in fact more general than the lowest-order expansion we consider here, since it depends only on the mechanism by which the critical point disappears at $n_c$.\cite{baxter latent heat, cardy nauenberg scalapino, Newman Riedel Muto} 

\begin{figure}[t] 
\centering
\includegraphics[height=1.in]{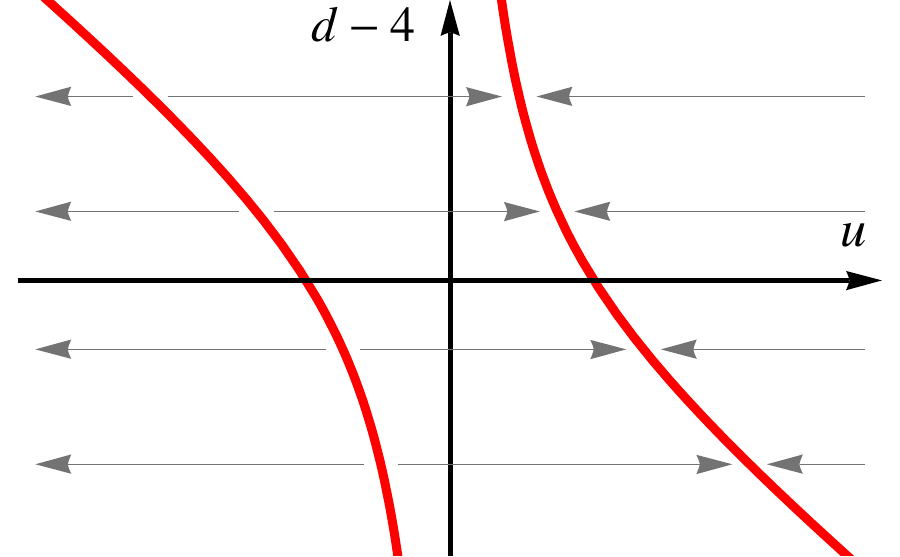} \hspace{0.1mm}
\includegraphics[height=1.in]{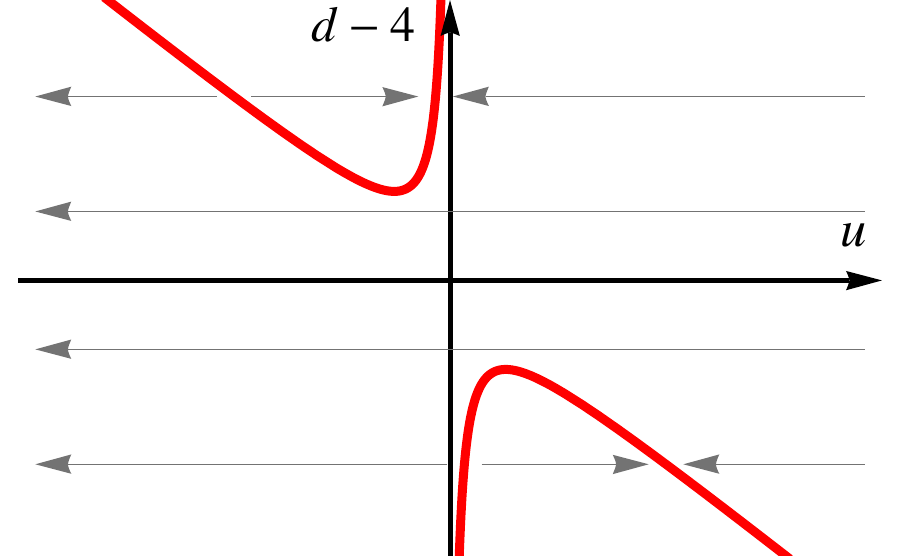} 
\caption{(color online) RG fixed points and flows in the $(u,d)$ plane. Left: for $n<2$. Right: for $n>2$.}
\label{RGflowsudplane}
\end{figure}

In Fig.~\ref{RGflowsudplane} we show the RG fixed points in the $(d,u)$ plane for $n>2$ and $n<2$. Note that when $n<2$ the fixed points below four dimensions are smoothly connected to those above. This is in agreement with our belief that a $6-\epsilon$ expansion is possible in the model with $n=1$.\cite{vortex paper} For $n>2$ a branch of fixed points appears above four dimensions. These fixed points, which are at negative $u$, are not expected to correspond to genuine critical points, as a result of unboundedness of the fixed point potential (this phenomenon is present even in the $O(N)$ model \cite{Wilson Fisher}).

\subsubsection{More concrete picture}
\label{more concrete picture}

Let us rewrite the soft spin $CP^{n-1}$ model as
\be
\mathcal{L}_\text{soft spin} = \tr (\nabla Q)^2 + t \tr Q^2 + \f{u}{22} \tr Q^4 + V(Q)\,,
\ee
collecting the operators that vanish at $n=2$ into $V(Q)$. There is one of these at cubic and one at quartic order, and we allow for higher terms with
\be
V(Q) = g_1 \tr Q^3 + g_2 \lf \tr Q^4 - \f{1}{2} (\tr Q^2)^2 \ri+\ldots
\ee
The two operators shown explicitly vanish when $n=2$, by virtue of the tracelessness of $Q$.

The RG equations for $t$ and $u$ must of course be independent of the $g_i$ when $n=2$, so any term which depends on the latter must have a coefficient  proportional to $\Delta$. This phenomenon  also occurs in the Potts model.\cite{Newman Riedel Muto} To the order that we require, the RG equations for $u$ and $g_i$ hence have the form
\ba \notag
\f{\dd \uresc}{\dd \ln L} & = \epsilon \uresc - \uresc^2 - \Delta \, f (g_1,g_2,\ldots), \\
\f{\dd g_i}{\dd \ln L} & = \beta_i(g_1,g_2,\ldots), \label{Full RG equations}
\end{align}
and we have checked explicitly that the $f$ term is of order $\Delta$, not higher. The cubic coupling $g_1$ is strongly relevant at the four-dimensional Gaussian fixed point, which obstructs a perturbative calculation of $f$ and $\beta_i$. However to obtain Eq.~(\ref{modified RG equation}) we need only assume that the $g_i$ flow to fixed point values $g_i^*$ under the RG equations (\ref{Full RG equations}), and expand around these with $g_i = g_i^* + \delta g_i$, obtaining
\ba \notag
\f{\dd \uresc}{\dd \ln L} & \simeq \epsilon \uresc - \uresc^2 - \Delta f(g_1^*, g_2^*,\ldots), &
\f{\dd \delta g_i}{\dd \ln L} & \simeq - b_{ij} \delta g_j.
\end{align}
The first of these equations yields Eq.~(\ref{modified RG equation}), with $a=f(g_1^*, g_2^*,\ldots)$. The second yields subleading (order one) irrelevant exponents  associated with the operators in $V(Q)$.

\subsubsection{Two upper critical dimensions at $n=2$}
\label{UCDs}

In order to analytically continue in $n$ it is important to realise that it is the operators in $V(Q)$, and not their couplings, which vanish when $n=2$. We may consider higher-dimensional versions of the loop models discussed here, and this observation leads to the conclusion that for $n=2$ such models have two distinct upper critical dimensions.

For correlators that can be written down using only two spin indices, we do not need a replica limit or SUSY. We set $n=2$ directly, giving the $O(3)$ model with upper critical dimension four.  At the critical point these correlators will thus have Gaussian behaviour for $d\geq 4$. However for correlators that require more than two indices, we are forced either to analytically continue in $n$ or to use SUSY. In either case, the cubic term reappears (in the SUSY formulation, the soft-spin $\cp^{n+k-1|k}$ model, it is $\str Q^3$, where $\str$ is the supertrace), leading to a non-Gaussian theory below six dimensions. These correlators are thus expected to have nontrivial behaviour for $d<6$.   This could be tested numerically by a simulation in  four dimensions.



\section{Simulations}\label{simulations}

\subsection{Monte Carlo procedure}\label{MC}

To describe our Monte Carlo procedure, we explain in the following how configurations are labelled,
how an initial state is constructed, and what Monte Carlo updates are used. Our approach is similar to ones used in loop algorithms for simulations of quantum spin systems.\cite{QMCreviews}
 
We generate the fugacity $n$ from a sum on loop colours, and so a configuration
of the model is specified by the pairing of incoming and outgoing links at each node and by
a colour for each link, with the constraint that all links belonging to the same loop
must have the same colour. This approach imposes the requirement that $n$ is integer:
we have developed an alternative algorithm that works for any real positive $n$, but 
it is less efficient.

An initial state is constructed by choosing at random the configuration of each
node with the specified probability. A colour is associated with each loop, chosen with equal probability from $n$ alternatives.

Subsequent states are generated using three kinds of Monte Carlo move. In the first, a node is chosen at random.
If the two loop strands passing through the chosen node have
different colours, the node configuration is not changed, but if
they both have the same colour, it is changed
according to the following rules. 
Denote the node configuration that has
probability $1-p$ by $\alpha$ and the one with probability $p$ by $\beta$.  
For $p<1/2$ a node
in configuration $\alpha$, is changed to $\beta$ with probability $p/(1-p)$, and one in configuration 
$\beta$ is always changed to $\alpha$. For $p>1/2$, a node in configuration 
$\alpha$ is always changed to $\beta$, and a node in configuration $\beta$ is changed 
to $\alpha$ with probability $(1-p)/p$. 
In the second type of Monte Carlo move, a link is chosen at random and the colour of all 
the links of the loop to which it belongs is changed to a different colour,
chosen with uniform probability from the $n-1$ possibilities.
The third type of move is to re-colour all loops in the system, with the new colours
selected independently and at random for each loop. It is designed to ensure that the colours of
short loops equilibrate efficiently.

The first two types of move are intercalated, with ten node updates followed by one colour change. 
For a sample with $N$ nodes we call $N/10$ such sequences a Monte Carlo sweep.
Measurements are performed every two Monte Carlo sweeps and the third type of move
is applied after each measurement sweep. The autocorrelation function of the energy is 
used to estimate a correlation time.

We consider loop models with an integer number of colours $1\leq n\leq 10$,  and system sizes of up to $7.5\times10^5$ links for $n\geq2$, and $6\times10^9$ for $n=1$.
The minimum number of Monte Carlo sweeps used is $10^5$ for any $n$, $p$ and $L$, and increases with decreasing $L$.

\subsection{Observables}

We measure observables for the loop models that are related to those of the $CP^{n-1}$ model.
In particular, we calculate quantities with the same scaling behaviour as the stiffness, susceptibility, order parameter and heat capacity of the
sigma model, and we compute the Binder cumulant for the
energy. We also evaluate the fractal dimension of loops. Detailed definitions and expected
finite-size scaling behaviour are as follows.

\subsubsection{Stiffness}

As a quantity equivalent to 
the sigma model stiffness, we study the average
number $n_{\rm w}(p,L)$ of curves winding around the sample
in a given direction.\cite{stiffness} More precisely, we pick a plane of
the lattice and count for each configuration the number
of sections of trajectory that leave this plane on a given
side and wind around the sample to reach the same plane
from the opposite side.
It is a property of the models that for each such trajectory section winding in the given
direction, there is another trajectory section winding in the opposite direction.  
For large sample size $n_{\rm w}(p,L)$ approaches zero in 
a phase with only short loops, and is proportional to $L$
in a phase with extended trajectories.
If there is a continuous transition between these phases at a critical point $p_{\rm c}$
with correlation length exponent $\nu$,
one expects the finite-size scaling behaviour
\begin{equation}
 n_{\rm w}(p,L) =f_{\rm w}(L^{1/\nu}[p-p_{\rm c}])\;.\label{scaling}
\end{equation}

\subsubsection{Susceptibility}

The susceptibility of the sigma model can be expressed in the standard way as the spatial integral
of the connected part of a two-point correlation function. With this as motivation,
let $n(s)$ be the number of loops of length $s$ in a given configuration. Then the spatial integral of the two-point correlation function in Eq.~(\ref{15}) is proportional to $\langle {\sum}_s s^2 n(s) \rangle$, where $\langle\ldots\rangle$ denotes an average over configurations. We split $n(s)$ into contributions $n_\text{ext}(s)$ and  $n_\text{loc}(s)$ from extended and localised loops, with $n(s) = n_\text{loc}(s) + n_\text{ext} (s)$. Here we define extended loops to be those that contribute to $n_\mathrm{w}$. We then take as our definition of the susceptibility $\chi$
\begin{eqnarray}
\chi L^3&=&\left\langle\sum_{s=0}^{L^3}s^2n(s)\right\rangle
-\left\langle\sum_{s=0}^{L^3}s^2n_{\rm ext}(s)\right\rangle\label{sus}\nonumber \\
&=&\left\langle\sum_{s=0}^{L^3}s^2n_{\rm loc}(s)\right\rangle\,.
\end{eqnarray}
On approaching a continuous transition, $\chi$ in an infinite system diverges 
with a critical exponent $\gamma$, while the expected scaling form in a finite system is
\begin{equation}
\chi=L^{\gamma/\nu}f_\chi(L^{1/\nu} [p-p_c])\;.
\end{equation}

\subsubsection{Order parameter}

The value of the order parameter $\cal M$ can be extracted from the correlation function
used to compute the susceptibility: the disconnected part -- the second term on the right hand side of
Eq.~(\ref{sus}) -- is proportional to ${\cal M}^2$. Hence we take
\begin{equation}
{\cal M} L^3=\sqrt{\left\langle\sum_{s=0}^{L^3}s^2n_{\rm ext}(s)\right\rangle}\;.
\end{equation}
At a continuous transition  $\cal M$ varies with critical exponent $\beta$
and has the finite-size scaling behaviour
\begin{equation}
{\cal M}=L^{-\beta/\nu}f_{\cal M}(L^{1/\nu} [p-p_c])\;.
\end{equation}

\subsubsection{Heat capacity}

The heat capacity $C$ can be expressed in the usual way in terms of fluctuations in the energy.
Absorbing a constant, we take from Eq.~(\ref{energy})
\begin{equation}
CL^3=\langle N_{p}^2\rangle -\langle N_{p}\rangle^2 \,.
\end{equation}
Introducing the critical exponent $\alpha$, we expect the finite-size scaling form
\begin{equation}
C=L^{-\alpha/\nu}f_C(L^{1/\nu} (p-p_c))\;.
\end{equation}

\subsubsection{Fractal dimension}

To evaluate the fractal dimension $d_f$ of loops we measure the end-to-end distance $R(s)$ of portions of trajectories as a function of arc-length $s$. Evidently, while on average $R(s)$ increases with $s$ for small $s$, it must decrease to zero for each loop as $s$ approaches the loop length.
To eliminate these finite-loop effects we retain only those contributions to $R(s)$ for which $s$
is less than one-third of the loop length. 
We then expect
\begin{equation}
\langle R(s) \rangle\propto s^{1/d_f}\,.
\end{equation}

\subsubsection{Binder cumulant}

As a tool for distinguishing between first order and continuous transitions, we compute the
Binder cumulant $V_L$ for the energy. 
This is defined by
\begin{equation}
V_L\equiv 1-\frac{1}{3} \frac{\left\langle n_p^4\right\rangle}{\left\langle n_p^2\right\rangle^2}\;.
\end{equation}

%

\subsection{Critical behaviour for $CP^1$}\label{cp1}


We use the loop model with $n=2$ as a test case, since there is a clear expectation that
it should have a phase transition with critical behaviour in the same universality class as
that of the $O(3)$ model in three dimensions, which in turn 
is known to high precision from
previous simulations.\cite{O(3) numerics} We study system sizes $32\leq L \leq 100$. We find similar results on 
both the K-lattice and the L-lattice. For conciseness we present data only in the former case.

\begin{figure}[!h] 
\centering
\includegraphics[height=2.2in]{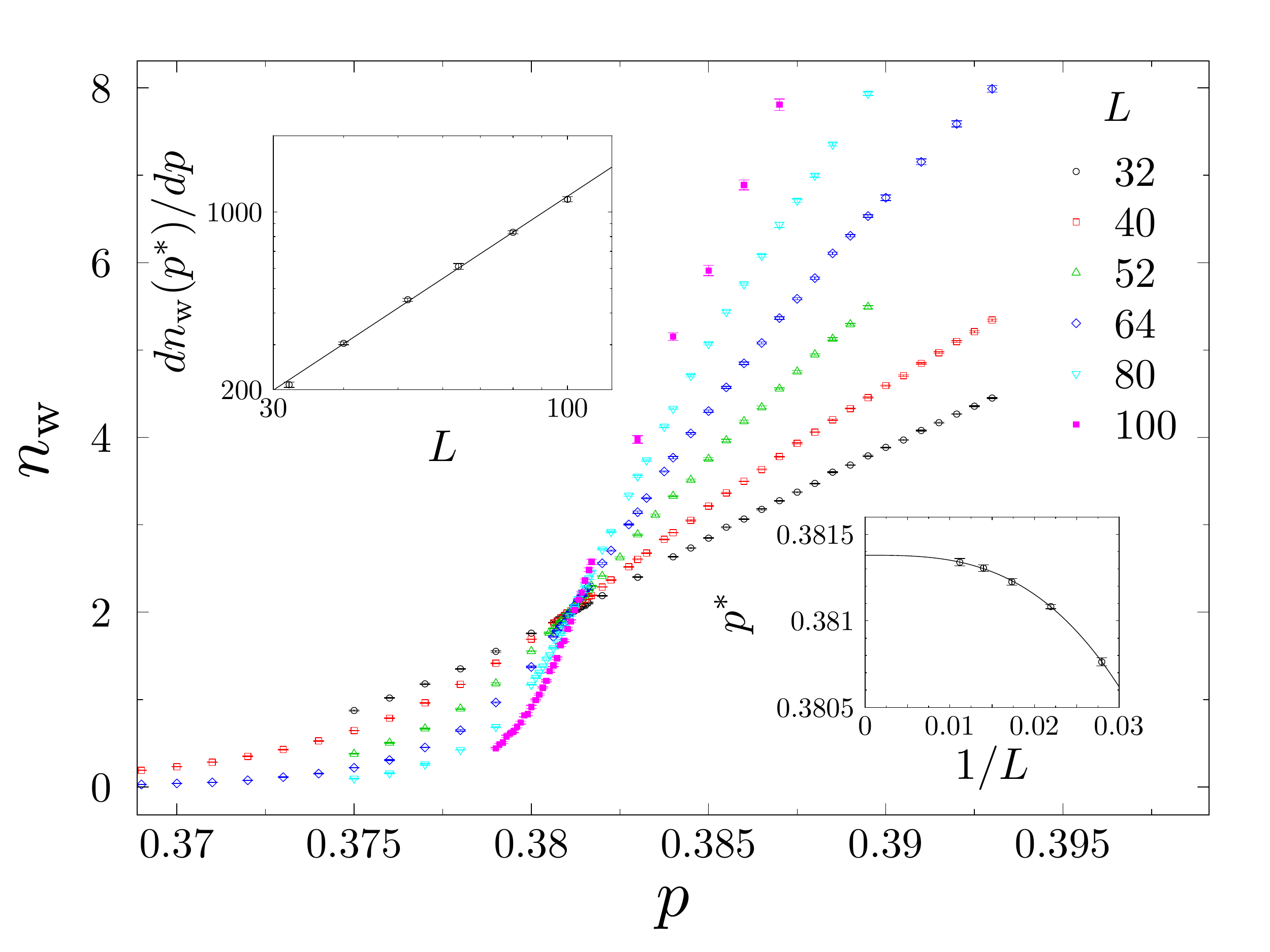} 
\caption{(color online) Winding number for $n=2$. Main panel: $n_{\rm w}(p,L)$ as a function of $p$ for different system sizes $L$. Left inset: ${\rm d}n_{\rm w}(p)/{\rm d}p$ at crossing points $p^*$ vs. $L$ on log-log scales. Right inset: $p^*$ vs $1/L$.}
\label{n_W at n=2}
\end{figure}

The existence of a phase transition is evident from the behaviour of $n_{\rm w}(p,L)$
displayed in Fig.~\ref{n_W at n=2}. 
The winding number is expected to decrease with increasing system size 
in a phase with only short loops,
and to increase with system size in a phase with extended loops. 
In confirmation, curves of $n_{\rm w}(p,L)$  as a function of $p$ cross at a common point 
for different $L$. To illustrate this in detail, we show in
the lower right inset to Fig.~\ref{n_W at n=2} the crossing point $p^*$ for curves
at two successive system sizes $L_1$ and $L_2$ as a function of 
the inverse of the geometrical mean size $L=\sqrt{L_1 L_2}$. We fit this to the form 
$p_{\rm c}+a/L^b$ (full line), obtaining
$p_{\rm c}=0.38138(4)$ and $b=3.0(5)$. 
For a first approach to determining the exponent $\nu$, we plot in the upper left
inset to Fig.~\ref{n_W at n=2} the gradient $dn_{\rm w}(p^*)/dp$ as a function of
$L$ on a double logarithmic scale. As expected from Eq.~(\ref{scaling}), the data
fit well to a straight line. The inverse gradient of this line yields the estimate $\nu = 0.68(3)$.
 
An alternative method for the determination of $\nu$ is to attempt
a scaling collapse of all data. In this approach we construct the scaling
function numerically, allowing for a non-linear dependence of the scaling variable on
distance from the critical point and including corrections to scaling governed by a leading
irrelevant exponent $y_{\rm irr}<0$. The quality of the fit and the number of 
fitting parameters supported by the data are judged by the value of $\chi^2$
compared to the number of degrees of freedom.
We use for the scaling variable
\begin{equation}
x= L^{1/\nu} \left[(p-p_{\rm c}) + A(p-p_{\rm c})^2 \right]\,,
\label{x variable}\end{equation}
finding that inclusion of the term in $(p-p_{\rm c})^2$ is justified by the fit, while
addition of a further term proportional to $(p-p_{\rm c})^3$ would not be.
We construct a fitting function 
$f_{\rm splines}(x)$ using cubic B-splines with 16 points. We
combine it with corrections to scaling, characterised by $y_{\rm irr}$ and
an $m$th order polynomial $P_m(x)$, in two alternative ways: either as
\begin{equation}
f_1(x,L)=f_{\rm splines}(x)[1+P_m(x)L^{y_{\rm irr}}]
\end{equation}
or as
\begin{equation}\label{f2}
f_2(x,L)=f_{\rm splines}(x)+P_m(x)L^{y_{\rm irr}}\,.
\end{equation}
We achieve the best fit using the form $f_1(x,L)$ with $m=1$. It yields
$\nu=0.706(8)$, $y_{\rm irr}=-1.0(3)$ and $p_{\rm c} = 0.38141(5)$ with
$\chi^2=300.5$ for 291 degrees of freedom. This scaling collapse is illustrated in Fig.~4 of
Ref.~\onlinecite{prl}.

\begin{figure}[!h] 
\centering
\includegraphics[height=2.2in]{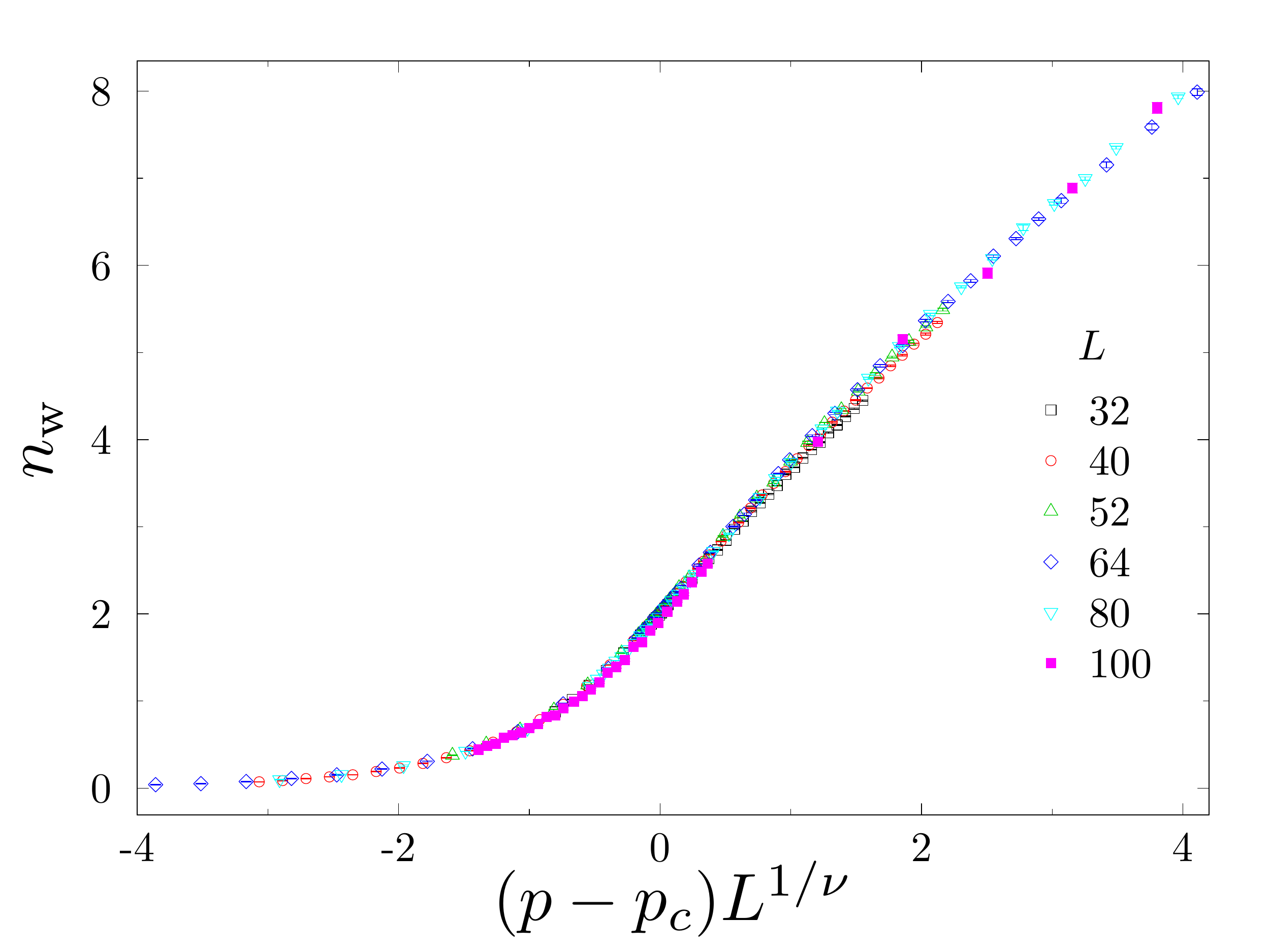} 
\caption{(color online) Scaling collapse at $n=2$ of $n_{\rm w}(p)$ as a function of $L^{1/\nu}(p-p_{\rm c})$ using the best estimate of  $\nu$ for the $O(3)$ model, without allowing for corrections to scaling.}
\label{n_W collapse at n=2}
\end{figure}

As a further demonstration that our results for $n=2$ are indeed compatible 
with the universality class of the $O(3)$ model, we attempt a scaling collapse
using the best available exponent estimate\cite{O(3) numerics} for that class, $\nu=0.7112$.
We omit finite size corrections, leaving the value of $p_{\rm c}$ as the only 
fitting parameter. The fitted value of $p_{\rm c}$ is $0.38120(10)$. This procedure results in good overlap of data from different systems sizes,
as illustrated in Fig.~\ref{n_W collapse at n=2}.

We analyse data for the susceptibility and for the order parameter by adapting
the approach summarised for the winding number in Eqns.~(\ref{x variable}) - (\ref{f2}).
Taking into account the fits for all three observables, our best exponent estimates are
$\nu=0.708(5)$ and $\gamma=1.39(1)$. 

\begin{figure}[!h] 
\centering
\includegraphics[height=2.2in]{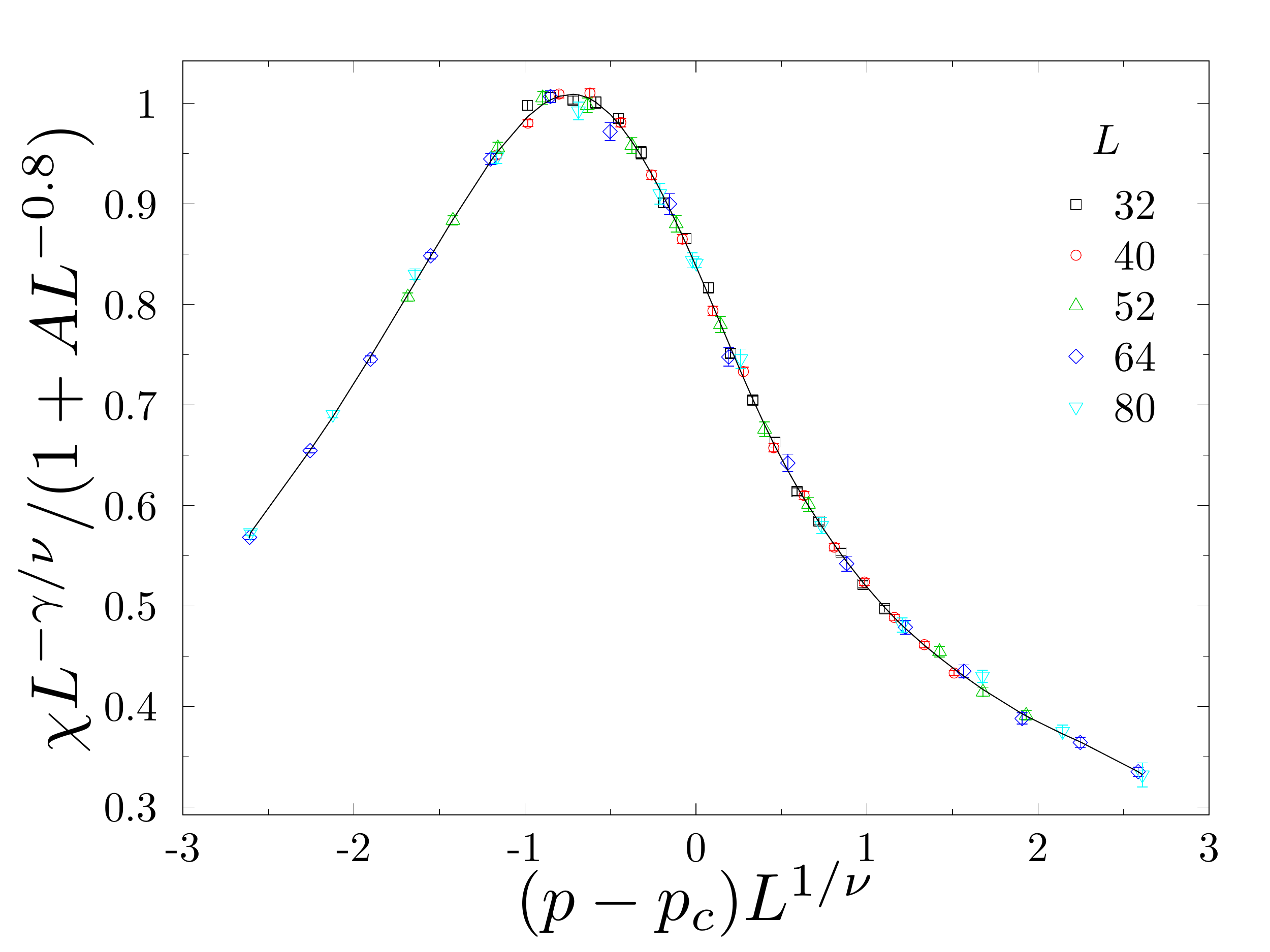} 
\caption{(color online) Scaling collapse of data for susceptibility $\chi$ as a function of $(p-p_{\rm c})L^{1/\nu}$ at $n=2$ using for the exponents the best estimates for the $O(3)$ model.}
\label{chi collapse at n=2}
\end{figure}

As an alternative, we also attempt collapse of data for the susceptibility 
using the best exponent estimates for the $O(3)$ model. The outcome is
shown in Fig.~\ref{chi collapse at n=2}: 
we plot $\chi L^{-\gamma/\nu}/(1+AL^{y_{\rm irr}})$ as a function of
$L^{1/\nu}(p-p_{\rm c})$, fixing the values \cite{O(3) numerics}
$\nu=0.7112$, $\gamma=1.3960$ and $y_{\rm irr}=-0.8$.
This leaves the values of $p_{\rm c}$ and $A$ as the only fitting parameters:
the fitted values are $p_{\rm c}=0.38145(15)$ and $A=1.58(20)$.
Again we obtain good overlap of data from different system sizes.
We note that the values of $p_{\rm c}$ obtained using different methods are
consistent within statistical errors.

Measurement of the fractal dimension of loops at the critical point yields $d_{\rm f} = 2.475(20)$,
implying $\eta=0.05(4)$, which is consistent with the result $\eta=0.0375(5)$ from previous high-precision Monte Carlo studies.\cite{O(3) numerics} 

We have repeated the same procedure for data from the L-lattice with similar results.
From an analysis of the winding number, the susceptibility and the order parameter
on this lattice, we obtain for the critical exponents the values $\nu=0.708(4)$ and $\gamma=1.383(10)$. These values again agree within errors with the best estimates 
for the $O(3)$ model.\cite{O(3) numerics}

\subsection{Identifying the order of transitions}


The phase transition in the loop model on the K-lattice may be continuous or first order, 
depending on the value of $n$, and in this subsection we set out our approaches
to determining the order of the transition from Monte Carlo data. 
Indeed, as discussed in Sec.~\ref{fourminusepsilon}, while we expect a continuous transition at $n=2$
from the equivalence to the $O(3)$ model, a first order transition 
would be natural for larger $n$ since Landau theory admits a cubic invariant if $n\not=2$.
Our results for the K-lattice are in fact consistent with a continuous transition at $n=3$ and with
first order transitions for $n\geq4$. Note that the loop model on the L-lattice
at $n\geq5$ has no extended phase but exhibits a first-order transition at $p=1/2$ between two 
symmetry-related short-loop phases (see Fig.~\ref{phase diagram}).

\begin{figure}[!h] 
\centering
\includegraphics[height=2.2in]{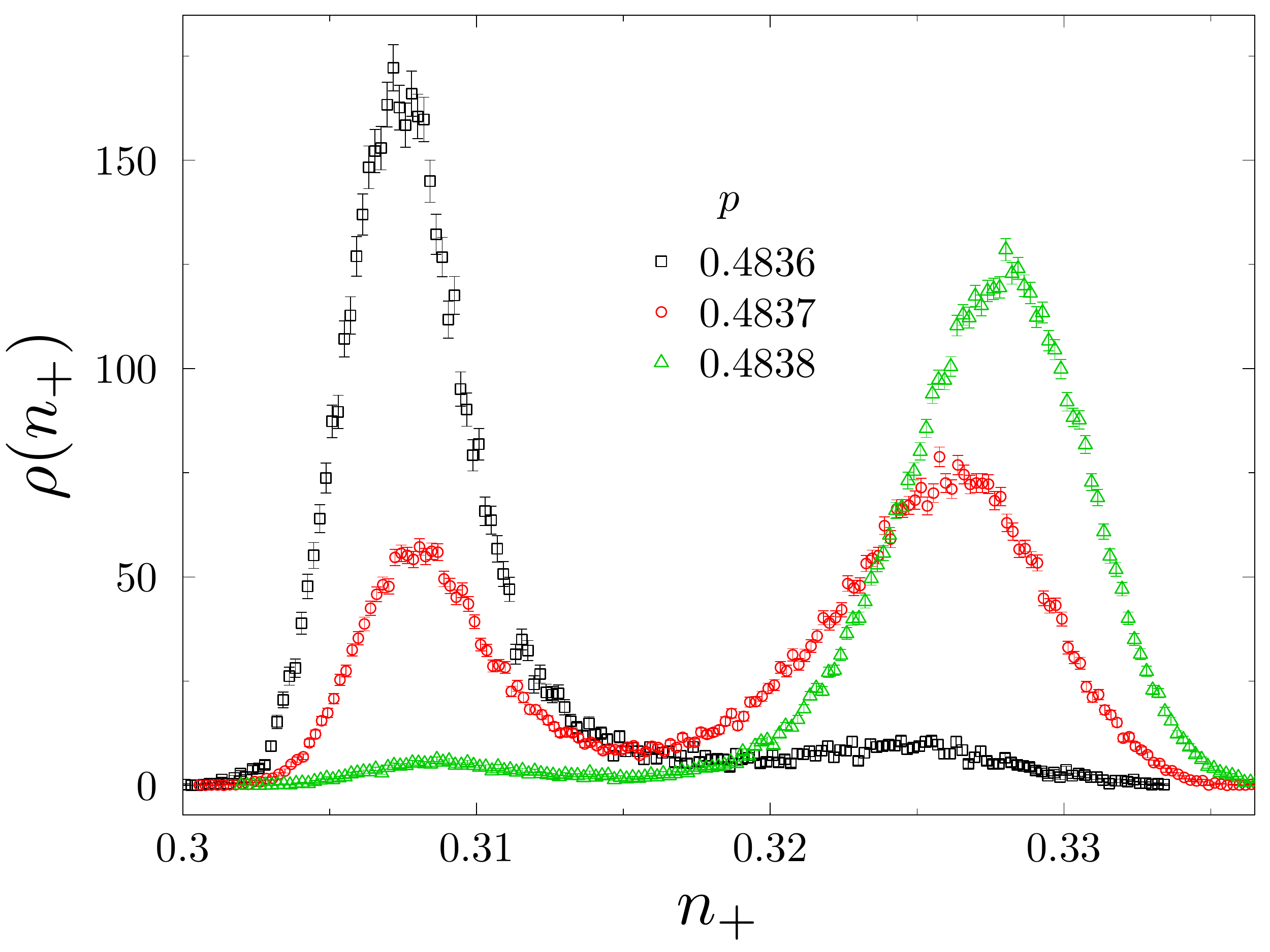} 
\caption{(color online) Evidence for a first order transition at $n=4$ on the K-lattice for $L=80$: probability distribution of the energy showing the double-peaked form that is characteristic of a first-order transition.}
\label{energy histogram at n=4}
\end{figure}

Distinguishing the order of a transition using simulations is delicate
in marginal cases because of finite size effects. We therefore begin by discussing
simple limiting examples. For $n=2$ we take it as established by the results of Sec.~\ref{cp1}
that the transitions on both lattices are continuous.
On the other hand, the transition for $n=4$ on the K lattice is  first order.
To demonstrate this, we show in Fig.~\ref{energy histogram at n=4}
the distribution of $n_+=N_p/(N_p+N_{1-p})$ (essentially the energy distribution) for
three different values of $p$ very close to the transition.
The double-peaked form, characteristic of a first-order transition, is very clear.

More generally, an established diagnostic for the order of a transition is provided by
the Binder cumulant $V_L$. In a system with a continuous transition
one expects $\lim_{L\to\infty} V_L = 2/3$ everywhere in the phase diagram, critical points included,
but at a first order transition point $\lim_{L\to\infty} V_L < 2/3$.\cite{binder}

\begin{figure}[!h] 
\centering
\includegraphics[height=2.2in]{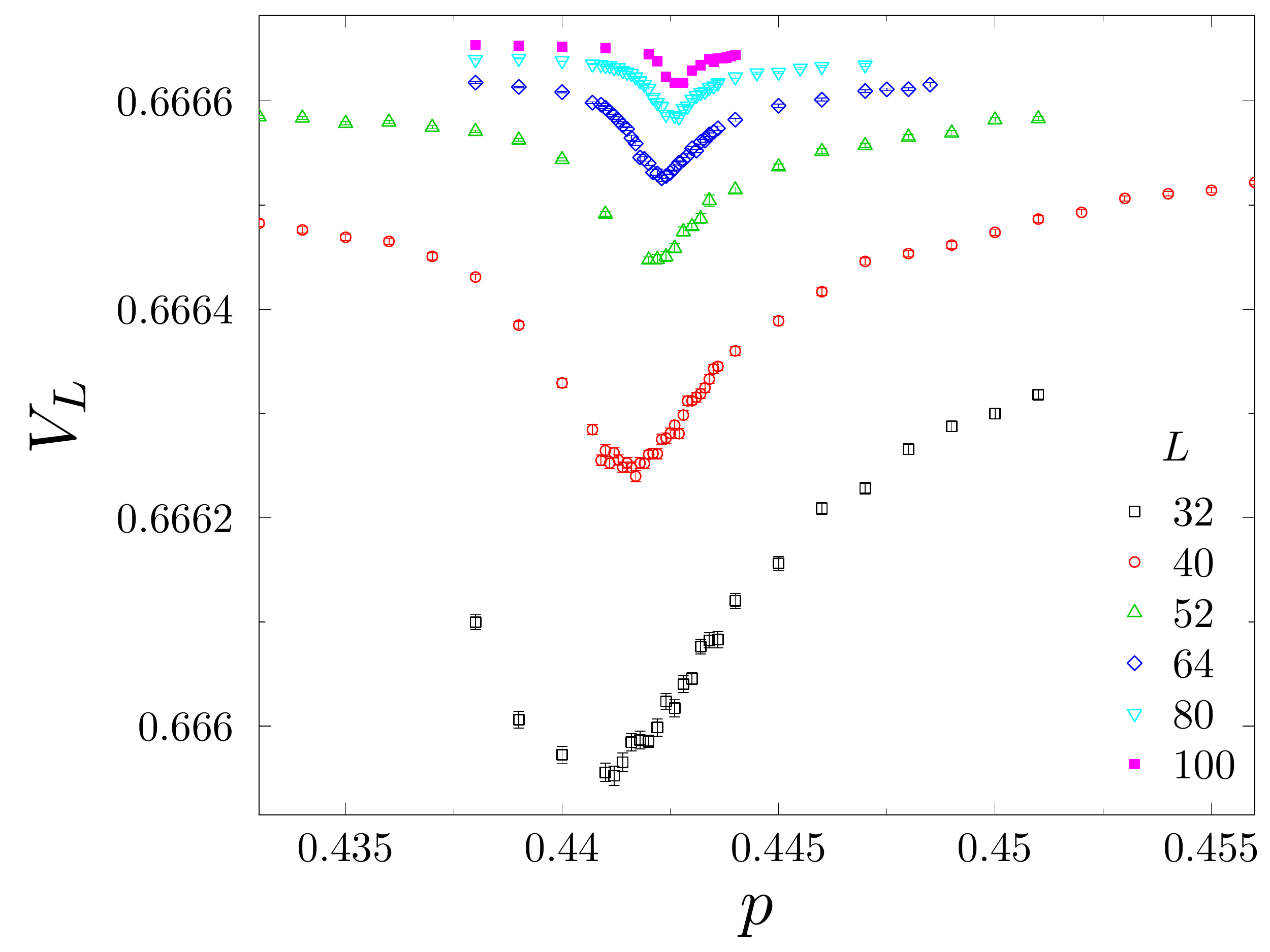} 
\caption{(color online) Binder cumulant $V_L$ at $n=3$: behaviour of $V_L$ as a function of $p$ for a sequence of system sizes $L$.}
\label{Binder cumulant at n=3}
\end{figure}

We illustrate behaviour of the Binder cumulant for $n=3$ on the K-lattice (anticipated to be a marginal case) in Fig.~\ref{Binder cumulant at n=3}. As a function of $p$, $V_L$ has a minimum near the transition point and approaches $V_L=2/3$ far from the transition on either side. The minimum becomes shallower with
increasing $L$ and the key issue is its limiting value. To focus on this we show in 
Fig.~\ref{[V_L]_{min} vs L} the difference $2/3-[V_L]_{\rm min}$ as a function of $L$ on a double logarithmic scale.
The data for $n=4$ on the K-lattice indicate a finite limiting value for $2/3-[V_L]_{\rm min}$
at large $L$, and hence a first order transition in this case. For other cases ($n=2$ and $n=3$ on both lattices, and $n=4$ on the L-lattice) the data fits a straight line with finite slope, as expected for
a continuous transition. If any of these transitions is in fact first order, the correlation length 
at the transition must be larger than 100 lattice spacings (but see Sec.~\ref{Latn=4} for a discussion of the L lattice transition with $n=4$).

\begin{figure}[!h] 
\centering
\includegraphics[height=2.2in]{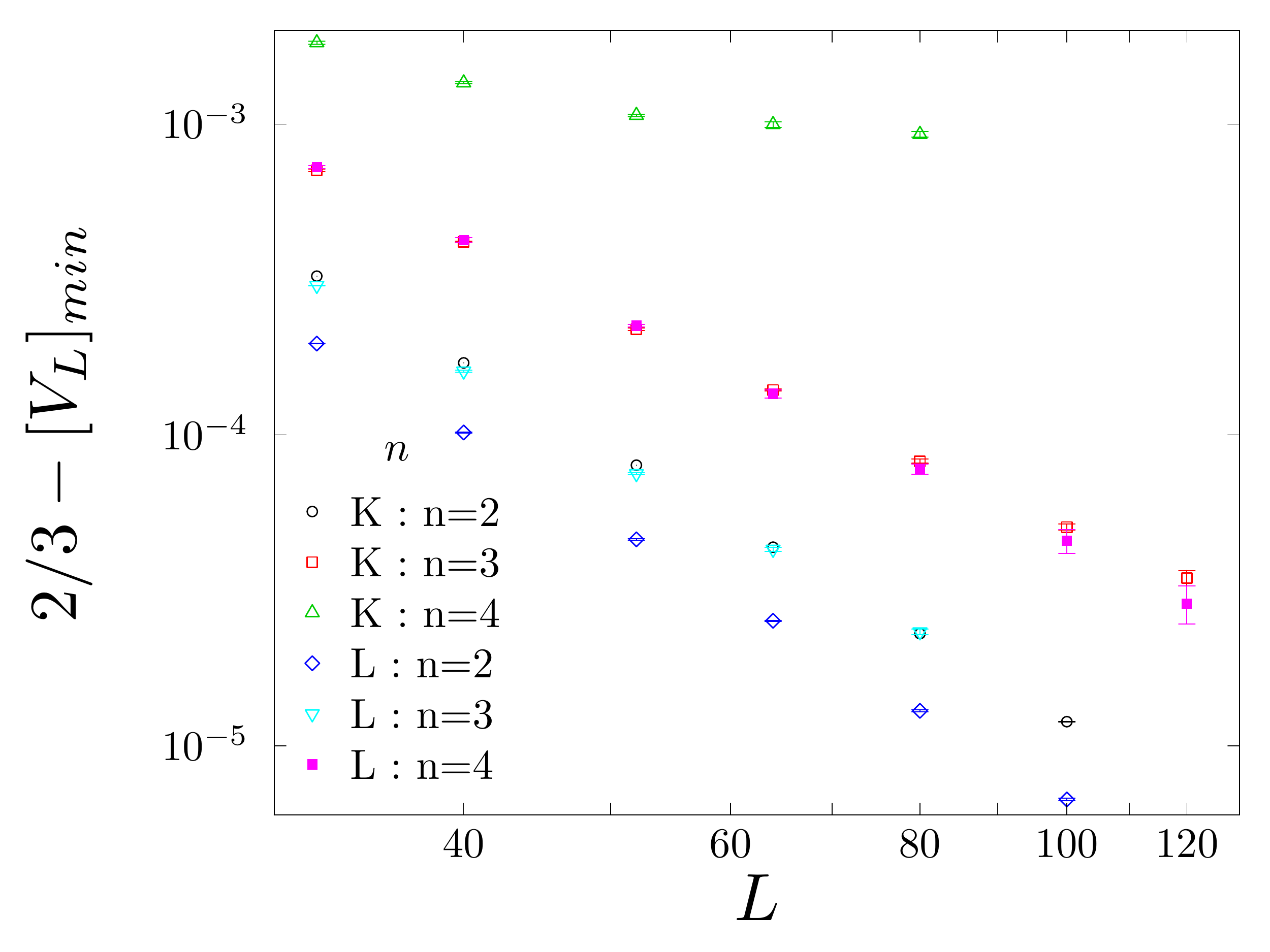} 
\caption{(color online) Distinguishing continuous from first order transitions: log-log plot of the dependence on system size of the deviation $2/3-[V_L]_{\rm min}$ of the minimum value of the Binder cumulant $V_L$ from its theoretical value far from the transition.  Data give evidence of a continuous transition at $n=2$ and $n=3$, and a first order transition on the K-lattice at $n=4$.}
\label{[V_L]_{min} vs L}
\end{figure}

\begin{figure}[!h] 
\centering
\includegraphics[height=2.2in]{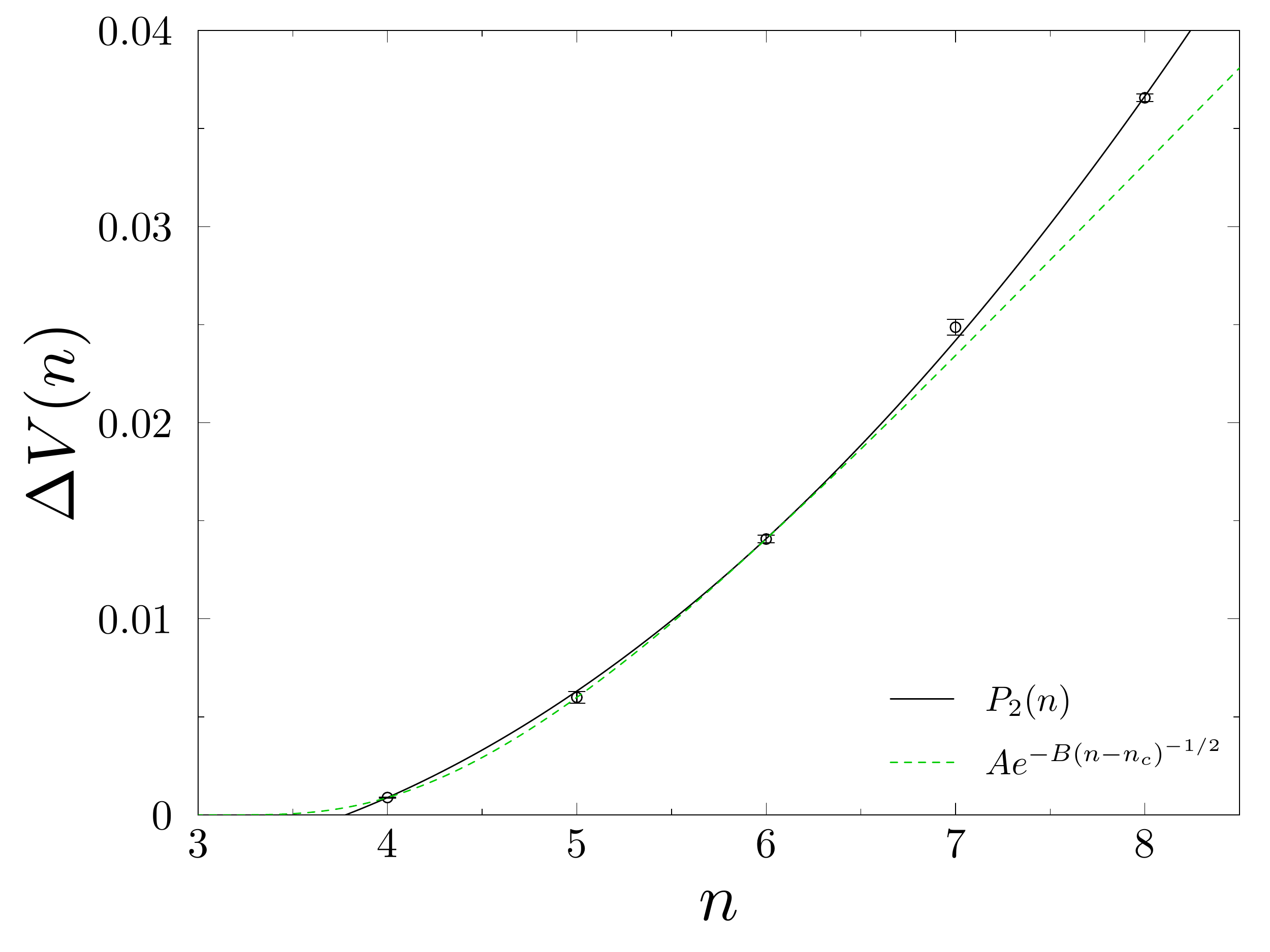} 
\caption{The discontinuity $\Delta V(n)$ at first order transitions as a function of $n$, with fits to a second-order polynomial $P_2(n)$ and to the functional form $Ae^{-B(n-n_c)^{-1/2}}$ 
motivated by Eq.~(\ref{ncrit}). The fitted values below which the transition is continuous
are $n_{\rm c} = 3.78(4)$ and $n_{\rm c}= 3.0(2)$, respectively.
}
\label{DeltaV(n)}
\end{figure}

A measure of the discontinuity at a first order transition is given by 
$\Delta V(n)=\lim_{L\to\infty} 2/3-[V_L]_{\rm min}$.
For the loop model with $n\ge 4$ on the K-lattice we can obtain 
$\Delta V(n)$ with high precision. As shown in Fig.~\ref{DeltaV(n)}, the dependence 
of  $\Delta V(n)$ on $n$
can be fitted to functional forms with a critical value $n_{\rm c}$ separating first order from continuous transitions that is larger than or close to 3. 
Thus the transition at $n=3$ is apparently second order, but $n_c<4$.

\begin{figure}[!h] 
\centering
\includegraphics[height=2.4in]{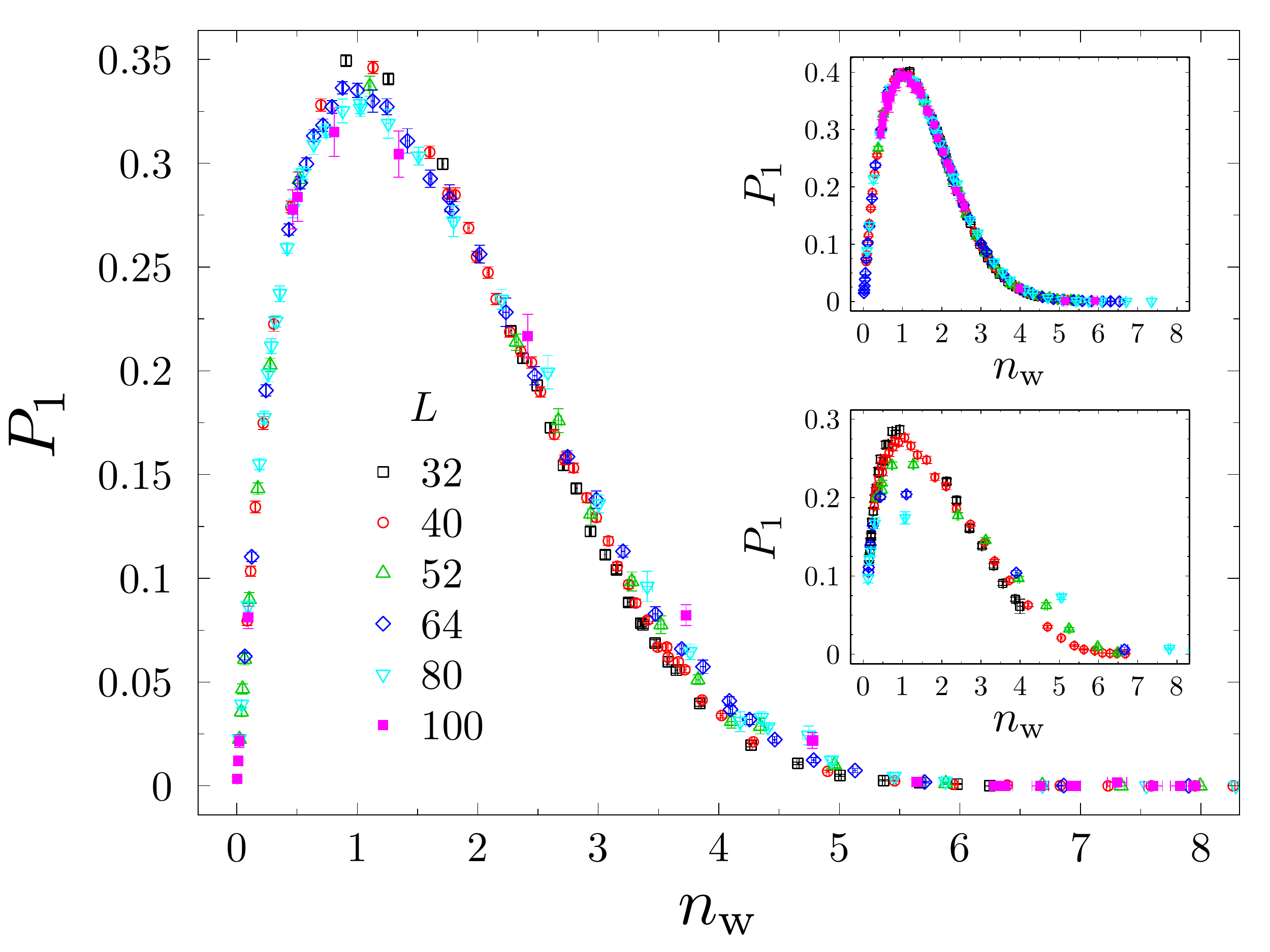} 
\caption{(color online) Parameter-free scaling collapse: data for the probability $P_1(p,L)$ that a configuration has exactly one winding curve as a function of the average number $n_{\rm W}(p,L)$ of winding curves: main panel $n=3$; top inset $n=2$. Bottom inset: contrasting behaviour at $n=4$.}
\label{P_1 vs n_W}
\end{figure}

An alternative method for characterising the nature of a transition is to
construct a parameter-free scaling function that has a fixed limiting form
if the transition is continuous, but not if it is first order. Consider for loop models
the probability $P_1(p,L)$ that a configuration has exactly one winding curve, as a function of the average number $\langle n_{\rm W}(p,L)\rangle$ of winding curves. 
At a continuous transition one expects this function to be independent of system size, 
provided finite-size corrections to scaling are not important. By contrast, 
the number of spanning curves jumps at a first order transition, from zero in one phase to a value proportional to $L$ in the other phase. In this case $P_1(p,L)$ therefore approaches zero
with increasing $L$  for all
$\langle n_{\rm W}(p,L)\rangle$.
In Fig.~\ref{P_1 vs n_W} we present this scaling function for $n=2$ (top inset), $n=3$ (main panel) and $n=4$ (bottom inset) for the K-lattice.
The data for different system sizes lie on a single curve for $n=2$, while those for $n=4$ do not.
Data for $n=3$ lie close to a single curve, although with larger deviations than at $n=2$.
Provided these deviations may be attributed to finite-size corrections to scaling, the
results of this analysis are consistent with the ones from the behaviour of the Binder cumulant.


\subsection{Critical behaviour for $CP^2$}

Having established that the transition in the loop model with $n=3$ on the K-lattice has a  
correlation length that appears divergent and certainly grows
larger than the largest accessible system sizes, we next characterise the critical behaviour.
We follow the methods described in Sec.~\ref{cp1} for $n=2$, except that in the
present case independent values of critical exponents are not available.

\begin{figure}[!h] 
\centering
\includegraphics[height=2.2in]{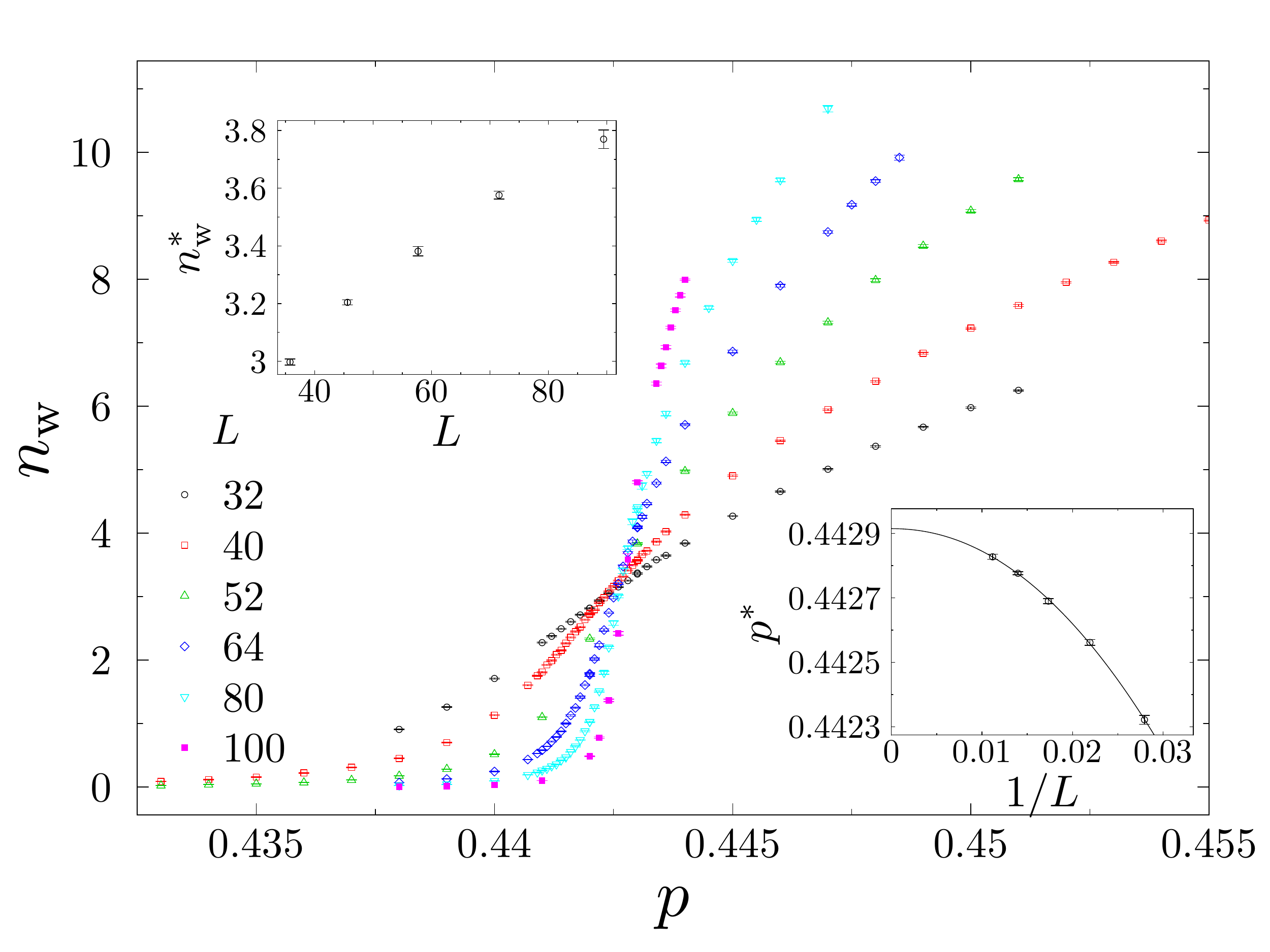} 
\caption{(color online) Scaling for $n=3$. Main panel: number of winding curves $n_{\rm w}(p)$ plotted as a function of $p$ for different system sizes $L$. 
Left inset: dependence of $n_{\rm w}$ at crossing points on $L$. 
Right  inset: $p^*$ vs $1/L$.}
\label{n_W at n=3}
\end{figure}

The unscaled data for $n_{\rm w}(p,L)$ are displayed
as a function of $p$ for several system sizes in Fig.~\ref{n_W at n=3} (main panel):
the transition point is apparent from the crossing of curves for different system sizes.
In the lower right 
inset to Fig.~\ref{n_W at n=3} we show the dependence of the probability $p^*$ at
which data for two successive sizes cross, as a function of the inverse of the geometrical
mean $L$ of these sizes. The continuous curve is a fit of the form $p_{\rm c}+a/L^b$,
with $p_{\rm c}=0.44291(2)$ and $b=2.1(3)$.
In the upper left inset to Fig.~\ref{n_W at n=3} we plot the dependence on $L$ of the winding number $n_{\rm w}$ at which data for two successive sizes cross. It shows that the finite size effects in this observable are not negligible. Nevertheless, we have undertaken a scaling collapse of all the data, using a non-linear scaling variable and including corrections to scaling as in Sec. V C. Results were presented previously in Fig. 4 of Ref.~\onlinecite{prl}. Improved statistics of data obtained more recently makes $\chi^2$ values for the collapse larger and seems to indicate that the finite size effects are not fully under control. The best estimates that we obtain are $\nu=0.51(2)$,
where the quoted error is statistical and systematic errors are difficult to estimate.

\begin{figure}[!h] 
\centering
\includegraphics[height=2.2in]{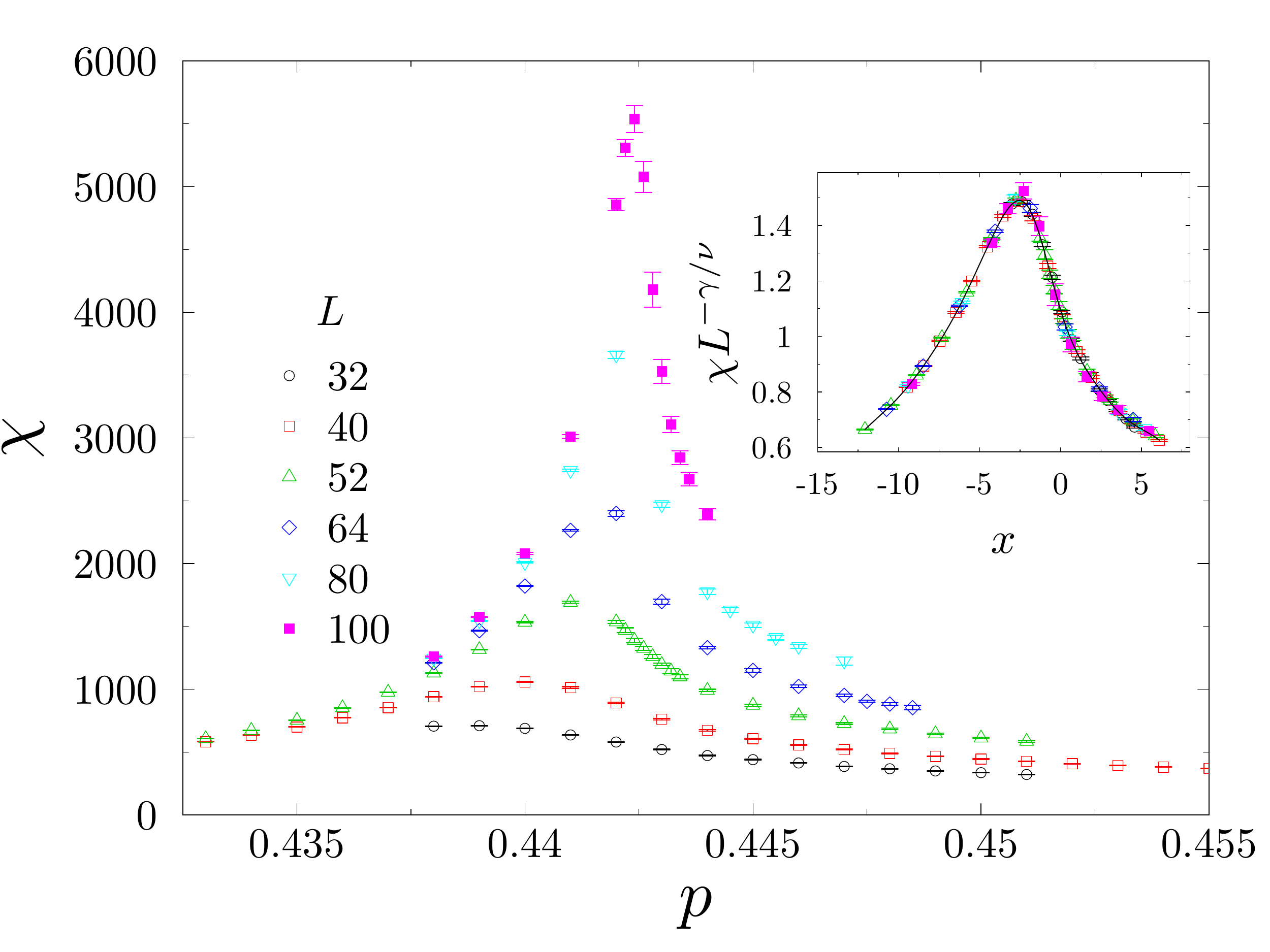} 
\caption{(color online) The susceptibility $\chi$ at $n=3$ for the K-lattice. Inset: scaling collapse of data in the main panel, using the scaling variable $x$ defined in Eq.~(\ref{x variable})}
\label{chi collapse at n=3}
\end{figure}

By contrast, we find that analysis of the scaling behaviour of the susceptibility $\chi$
is more straightforward. Its dependence on $p$ and system size is shown in 
Fig.~\ref{chi collapse at n=3}. It has a peak near the critical point which grows with $L$.
Plotting $\chi L^{-\gamma/\nu}$ as a function of 
$x=L^{1/\nu}[(p-p_{\rm c})+ A(p-p_{\rm c})^2]$,
these data exhibit scaling collapse, as illustrated in the inset to Fig.~\ref{chi collapse at n=3},
with $\chi^2=53$ for 60 degrees of freedom.
The exponent values resulting from this procedure are $\nu=0.542(16)$ and $\gamma/\nu=1.78(2)$, and $\gamma=0.97(2)$.

\begin{figure}[!h] 
\centering
\includegraphics[height=2.2in]{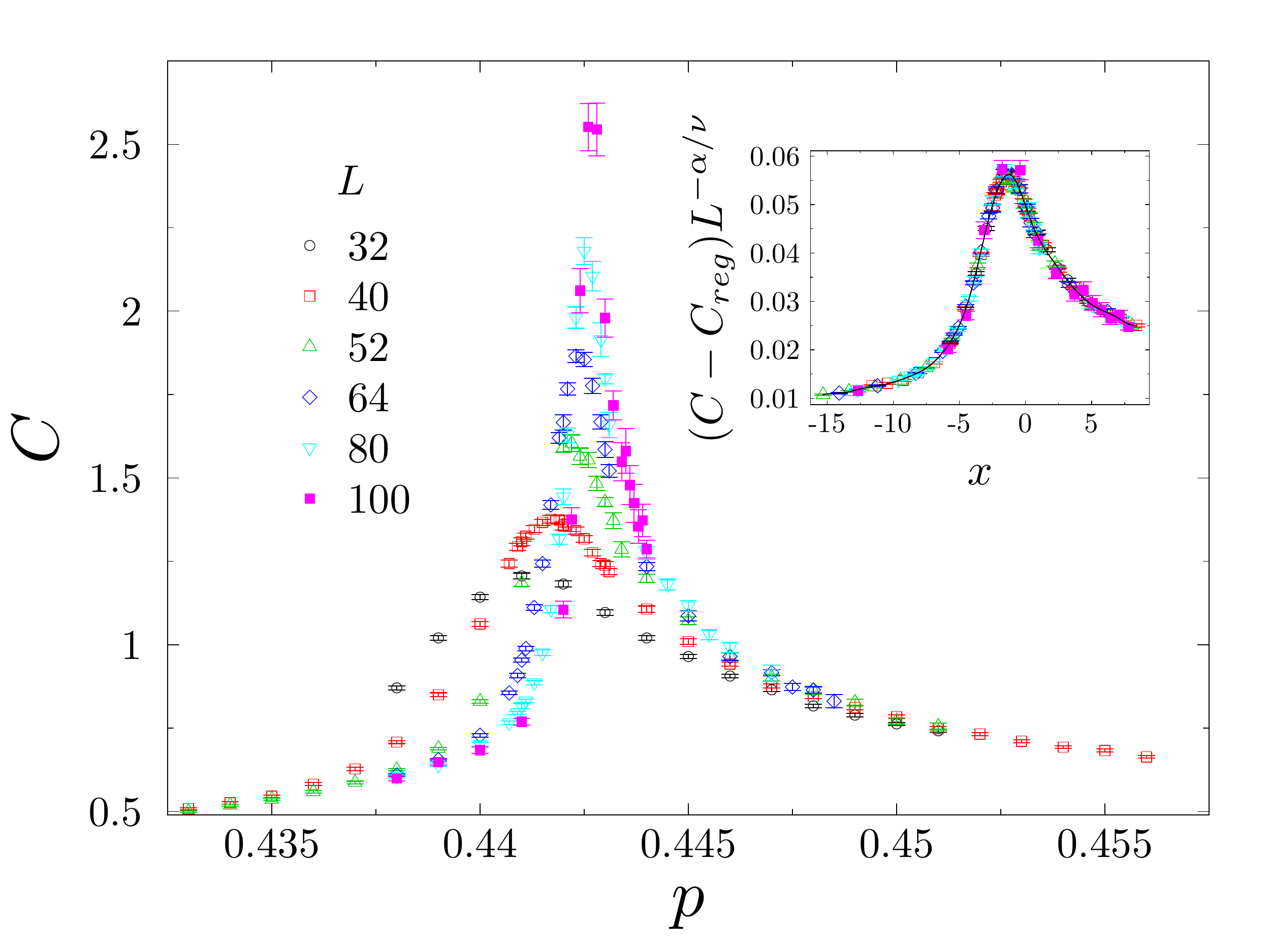} 
\caption{(color online) Heat capacity as a function of $p$ at $n=3$ for the K lattice. Inset: scaling collapse of the divergent contribution to the heat capacity.}
\label{C collapse at n=3}
\end{figure}

Data for the heat capacity $C$ are presented in 
Fig.~\ref{C collapse at n=3}. It is strongly divergent at the transition but
values in the smaller system sizes are dominated by a smooth background.
To effect a scaling collapse we therefore subtract an $L$-independent contribution
$C_{\rm reg}$ that varies smoothly with $p$, plotting $(C-C_{\rm reg})L^{-\alpha/\nu}$ 
as a function of  $L^{1/\nu}[(p-p_{\rm c})+A(p-p_{\rm c})^2]$: see Fig.~\ref{C collapse at n=3} inset.
From this approach we obtain $\chi^2=73$ with 95 degrees of freedom; assuming the $d$-dimensional hyperscaling relation $\alpha=2-d\nu$, we find $\nu=0.526(15)$. 

Inclusion of corrections to scaling
in the analysis of  $\chi$ and $C$ does not change significantly 
our exponent estimates or the scaling collapse, and we were unable to
determine a reliable value for $y_{\rm irr}$.

From a study of loops at the critical point (data not shown) we find the fractal dimension value
$d_f=2.40(3)$. This implies $\eta=0.20(6)$, which is broadly consistent with the value $\eta=0.22(2)$
obtained using hyperscaling and our analysis of $\chi$.


Combining the results from our analysis of data on the K-lattice for $\chi$, $C$, the order parameter (not shown), and using bootstrap methods to determine errors, our best estimates for the two independent critical exponents are $\nu=0.536(13)$ and $\eta=0.23(2)$. The scaling relations $\gamma=(2-\eta)\nu$ and $\beta=\nu(1+\eta)/2$ imply $\gamma=0.97(2)$ and  $\beta=0.33(1)$.
The given uncertainties are obtained from a purely statistical analysis. The results may also be affected by systematic errors: from the dispersion in the values of $\nu$ obtained from analysis of different
observables, we believe that the systematic errors in our exponent values are comparable in magnitude to the statistical ones.

Similar but less exhaustive analysis for the L-lattice gives compatible exponent values, although with
much larger uncertainties.

\subsection{L-lattice at $n=4$}
\label{Latn=4}

The case $n=4$ on the L-lattice deserves specific discussion. In the phase
diagram for this lattice (see Fig.~\ref{phase diagram}) the line $p=1/2$
is within an extended phase for small $n$ and is the location of first-order transitions for large $n$.
The point $p=1/2,\, n=n^*$ at which behaviour changes is expected to be a deconfined critical point.\cite{interacting} 
At $n=4$ as a function of $p$ we find two distinct transitions, at $p_c$ and $1-p_c$, with  $p_{\rm c}=0.4994(3)$, indicating that $n^*>4$. This is confirmed by the fact that behaviour at $p=1/2,\,n=4$ matches that expected in the extended phase:
$n_W(1/2,L)$ is quite accurately proportional to $L$ for our largest system sizes.
While we expect from universality and our results on the K lattice that the
transition on the L lattice at $n=4$ should be first order,
the proximity of the two transitions at $p_c$ and $1-p_c$ and of the
critical point at $n^*$ makes in natural that the correlation length at the transition should be very large.

\section{Concluding remarks}\label{conclusions}

In summary, we have shown that the loop models we consider provide lattice representations of 
$CP^{n-1}$ $\sigma$ models, and we have set out the correspondence between loop observables and $\sigma$ model correlators. The models have phase transitions between paramagnetic and ordered phases, which we argue using an RG treatment are continuous for $n\leq n_{c}$; from simulations we give evidence that $n_{c}>3$ in three dimensions. 

There is scope for further work on these and related loop models, in several directions. First, within the ordered phases of the models studied, a quantity of interest is the length distribution of long loops, expected to take a universal form.\cite{forthcoming length distribution paper}
Second, starting with the loop models on the L lattice at the symmetric point $p=1/2$, one can induce a phase transition by introducing an extra coupling. The short loop phase established in this way exhibits spontaneous symmetry breaking and the transition is a candidate for a deconfined critical point.\cite{interacting} Third, and separately, models with undirected loops are interesting as representations of $RP^{n-1}$ $\sigma$ models, corresponding at $n=2$ to the $O(2)$ model and at $n=3$ to models
for the  liquid crystal isotropic-nematic transition: it would be of interest to investigate the order
of the transition and possible critical behaviour as a function of $n$ for these undirected models.


\begin{acknowledgments}
We thank E. Bettelheim, P. Fendley, R. Kaul, I. Gruzberg, A. Ludwig, P. Wiegmann and especially J. Cardy for discussions. This work was supported by EPSRC Grant No.
EP/D050952/1, and by 
MINECO and FEDER  Grants No. FIS2012-38206 and AP2009-0668.
\end{acknowledgments}

\end{document}